\documentclass[aps,prl,reprint,showpacs,showkeys,longbibliography]{revtex4-2}

\usepackage[T1]{fontenc}
\usepackage[utf8]{inputenc}
\usepackage{amsmath,amssymb,amsfonts}
\usepackage{graphicx}
\usepackage{bm}
\usepackage{hyperref}   
\usepackage{orcidlink}

\hypersetup{
    colorlinks=true,
    linkcolor=blue,
    filecolor=magenta,
    urlcolor=blue,
    citecolor=blue
}

\usepackage{amsthm}

\theoremstyle{definition}

\newcommand{\PSL}{\operatorname{PSL}}

\usepackage{enumitem}
\newlist{assumption}{enumerate}{1}
\setlist[assumption,1]{
  label=\textbf{A\arabic*},
  ref=\textbf{A\arabic*},
  leftmargin=2em,
  itemsep=1ex plus 0.5ex
}

\begin{document}

\title{Geometric Origin of Exact Mean-Field Reductions: Möbius Symmetry and the Lorentzian Ansatz}
\author{Hugues Berry\orcidlink{0000-0003-3470-683X}}
\email{hugues.berry@inria.fr}
\affiliation{
AIstroSight, \href{https://ror.org/022gakr41}{Inria Lyon Centre}, \href{https://ror.org/01502ca60}{Hospices Civils de Lyon},\href{https://ror.org/029brtt94}{Université Claude Bernard Lyon 1}, 
F-69603 Villeurbanne, France
}

\author{Leonardo Trujillo\orcidlink{0000-0001-9995-4135}}
\email{leonardo.trujillo@insa-lyon.fr}
\altaffiliation[Permanent address: ]{
\href{https://ror.org/01rk35k63}{Université de Lyon}, \href{https://ror.org/050jn9y42}{INSA Lyon}, \href{https://ror.org/029brtt94}{Université Claude Bernard Lyon 1},  \href{https://ror.org/03p3f6k20}{CNRS UMR5240 MAP}, 69621 Villeurbanne, France
}
\affiliation{
AIstroSight, \href{https://ror.org/022gakr41}{Inria Lyon Centre}, \href{https://ror.org/01502ca60}{Hospices Civils de Lyon},\href{https://ror.org/029brtt94}{Université Claude Bernard Lyon 1}, 
F-69603 Villeurbanne, France
}

\date{\today}

\begin{abstract}
Low-dimensional descriptions of large systems of coupled oscillators and spiking neurons rely heavily on the Lorentzian Ansatz. We show that its privileged role is geometric rather than heuristic: for the transport induced by Riccati dynamics, the Cauchy-Lorentz family indeed emerges as the unique connected two-dimensional family of continuous probability densities that is invariant under the induced projective transport. The key step of the demonstration is to reformulate the dynamics on the circle, where the problem reduces to the uniqueness of the rotation-invariant probability measure. Under stereographic projection, this yields the standard Cauchy law and, under the full projective action, the Lorentzian family. This result gives a unified geometric foundation for the Ott-Antonsen [\href{https://pubs.aip.org/aip/cha/article-abstract/18/3/037113/341750/Low-dimensional-behavior-of-large-systems-of?redirectedFrom=fulltext}{ Chaos 18, 037113 (2008)}] and Montbri\'o-Paz\'o-Roxin [\href{https://doi.org/10.1103/PhysRevX.5.021028}{Phys. Rev. X 5, 021028 (2015)}] reductions, explains the failure of Gaussian closures, and identifies the structural condition underlying exact two-parameter reductions.
\end{abstract}

\maketitle

\paragraph{Introduction.---}
The search for low-dimensional macroscopic descriptions of high-dimensional complex systems is a central theme in statistical physics and nonlinear dynamics. A major step in this direction was the discovery by Watanabe and Strogatz (WS)\,\cite{WatanabeStrogatz1993,Watanabe1994} that large ensembles of globally coupled phase oscillators possess a remarkable integrable structure. Shortly thereafter, Goebel\,\cite{Goebel1995} identified that the WS transformation is structurally generated by Riccati dynamics and the associated M\"obius action. This geometric viewpoint was later developed by Marvel, Mirollo, and Strogatz (MMS)\,\cite{Marvel2009}, who showed that M\"obius symmetry provides the natural language for describing these finite-dimensional reductions. In the thermodynamic limit, this same structure underlies the exact continuum reductions of Ott and Antonsen (OA)\,\cite{ott2008low} for phase oscillators on the circle and of Montbri\'o, Paz\'o, and Roxin (MPR)\,\cite{montbrio2015macroscopic} for quadratic integrate-and-fire (QIF) neurons, a subtype of Riccati equations.

Central to these exact reductions is the assumption—often termed the \emph{Lorentzian Ansatz} (LA). In the MPR framework, the state density is directly assumed to remain within the Cauchy-Lorentz family, yielding an exact closure. The OA framework instead assumes a Poisson kernel for the state density on the unit circle $S^1$. These two assumptions are equivalent since the stereographic projection of the Poisson-kernel family on the circle $S^1$ is the Cauchy-Lorentz family on $\mathbb{R}$.
Therefore, under the Lorentzian ansatz, the infinite-dimensional population density collapses onto a finite-dimensional manifold parametrized by the center and width of the Lorentzian distribution, or equivalently by a complex order parameter. This raises a fundamental question: is the Lorentzian profile merely a convenient closure hypothesis, or is it selected by a deeper geometric principle? More generally, could other families, such as Gaussians, support exact reductions of the same kind? In the continuum setting, MMS made explicit that the family of Poisson kernels is invariant under a M\"obius  flow\cite{Marvel2009}. However, MMS does not address the  related uniqueness question--whether the Poisson family is the unique two-dimensional invariant family of continuous densities, or whether other families could exhibit the same invariance.

Recent work has further clarified the role of Riccati dynamics and M\"obius symmetry in finite-dimensional settings, including arrays of globally coupled Riccati units of arbitrary size\,\cite{cestnik2024ntegrability} and their complex generalizations\,\cite{pazo2025low}. However, in the physically relevant real-valued case and in the thermodynamic limit, the problem is no longer to describe individual trajectories but to classify invariant families of probability densities. 

In this Letter, we identify the geometric origin of the Lorentzian Ansatz. Our main result is that, for probability densities transported by real Riccati flows, the Cauchy-Lorentz family is not an arbitrary ansatz choice but actually the unique family of two-parameter continuous densities that is invariant. In other words, among all invariant families of continuous two-parameter densities, only the Lorentzian one is compatible with the M\"obius symmetry induced by the Riccati flow.

Our proof is geometric. After compactifying the real line to the circle, the classification problem reduces to identifying which continuous probability density on $S^1$ is compatible with the M\"obius action.
 The Riccati flow geometry belongs to the projective special linear group $\PSL(2,\mathbb R)$, of which one-parameter subgroups fall into three classes: elliptic, parabolic, and hyperbolic\,\cite{kisil2012geometry}. They act on $S^1$ respectively as rigid rotations, as flows with a single (double) fixed point, and as flows with two (simple) fixed points. We show that only the elliptic/rigid rotation class preserves a continuous, strictly positive density; the other two develop fixed points on $S^1$, which is incompatible with continuous invariance.
 Considering that the only continuous probability density that is invariant under a nontrivial rigid rotation on $S^1$ is the uniform (Haar) measure\,\cite{Halmos1950}, and that the stereographic projection of the Haar measure is the Lorentzian family, this establishes that the Cauchy-Lorentz family is the unique two-parameter family of continuous probability densities invariant under the Riccati flow.
Our result therefore bridges the dynamical perspective of Cestnik and Martens~\cite{cestnik2024ntegrability} with the classical rigidity of the Cauchy law under linear fractional transformations, and explains why OA/MPR closures are structural rather than ad hoc.

\paragraph{Geometric Origin of the Lorentzian Ansatz.---}
To understand why the Lorentzian profile is uniquely selected among all possible distributions, we look past the algebraic details of the Riccati equation and focus on the underlying geometry of the flow.

We begin our analysis at the microscopic level by considering a population of $N$ units whose states $z_j\in\mathbb R$ evolve according to the Riccati equation
\begin{equation}
\dot{z}_j = a(t)z_j^2 + b(t)z_j + c(t)
\end{equation}
with real coefficients. We recall in Supplemental Material {\color{red}S1}\,\cite{SupMat} that this nonlinear evolution can be linearized using the transformation $z_j=v_1/v_2$ (projective transformation) \,\cite{hille1976ordinary}. The dynamics of the vector $\mathbf{v}=(v_1,v_2)^\top\in\mathbb R^2$ is indeed given by a linear system
\[
\dot{\mathbf{v}}=\mathcal H(t)\mathbf v,
\qquad
\mathcal H(t)=
\begin{pmatrix}
b(t)/2 & c(t)\\
-a(t) & -b(t)/2
\end{pmatrix},
\]
where the choice of a trace-free matrix was made to fix the projectively irrelevant scalar gauge mode (see Supplemental Material {\color{red}S1}\,\cite{SupMat}). The time evolution is governed by the fundamental matrix $\mathbf M(t)$ solving $\dot{\mathbf M}=\mathcal H(t)\mathbf M$ or, alternatively $\mathbf v(t)=\mathbf M(t)\mathbf v(0)$. Note that $\mathcal H(t)\in\mathfrak{sl}(2,\mathbb R)$ (the Lie algebra of trace-free real $2\times 2$ matrices), so by Liouville's formula $\det \mathbf M(t)=1$; hence $\mathbf M(t)\in SL(2,\mathbb R)$.


Projecting back to the physical variable $z_j$ and writing
\[
\mathbf M(t)=
\begin{pmatrix}
\alpha(t) & \beta(t)\\
\gamma(t) & \delta(t)
\end{pmatrix},
\]
the state $z_j(t)$ evolves as
\begin{equation}
    z_j(t)=\Phi_t(z_j(0))
    =
    \frac{\alpha(t)z_j(0)+\beta(t)}{\gamma(t)z_j(0)+\delta(t)}.
    \label{eq:mobius_map}
\end{equation}
The rational function in the RHS term of eq.\eqref{eq:mobius_map} is, by definition, a Möbius transformation. Therefore, this shows that under Riccati dynamics, the states $z_j(t)$ evolve according to the Möbius flow $\Phi_t$. 

Because $z_j$ is defined as a ratio, multiplying $\mathbf M$ by a nonzero scalar does not change eq.\eqref{eq:mobius_map}; $\mathbf M$ therefore belongs to the projective group $\PSL(2,\mathbb R)=SL(2,\mathbb R)/\{\pm I\}$ (see Supplemental Material {\color{red}S1}\,\cite{SupMat}).

The dynamical laws derived here—the Riccati generator and the resulting Möbius flow—are structurally equivalent to Eqs.~(1) and (2) in Ref.\,\cite{cestnik2024ntegrability}. However, a crucial distinction must be made regarding the domain. While Cestnik and Martens consider the general complex case, here we restrict to the subgroup $\PSL(2,\mathbb R)$ acting on the real projective line. This restriction is not merely a simplification but a physical requirement: we are characterizing the evolution of a probability density over real observables such as membrane potentials. General complex Möbius transformations would map the real axis into circles in the complex plane, destroying the interpretation of $\rho(z)$ as a one-dimensional density of physical states.

In the thermodynamic limit $N\to\infty$, the empirical distribution of the $z_j$s converges to a probability density $\rho(z,t)$ obeying the continuity equation
\[
\partial_t\rho+\partial_z(\dot z\,\rho)=0.
\]
The exact solution of this equation is given by the push-forward of the initial density under the Möbius flow
\[
\rho(z,t)=(\Phi_t)_\sharp\rho(z,0),
\]
with $\rho(z,0)$ the initial density. Here $\Phi_\sharp\rho$ denotes the push-forward of $\rho$ by $\Phi$, i.e., the probability density obtained by transporting $\rho$ along the map $\Phi$, with normalization preserved\,\cite{villani2009optimal}.


We first ask the following question: if $\rho(z,0)$ is a Cauchy-Lorentz distribution, noted as $\rho(z,0)=\rho_{x,y}(z)=\frac{1}{\pi}\frac{y}{(z-x)^2+y^2}$, how does this initial density evolve? 

For convenience, we note the parameters of the initial density as a complex number $w=x+iy$ in the upper half-plane $\mathbb{H}=\{u+iv : v>0\}$, so that the initial density may be noted as $\rho_w(z)=\frac{1}{\pi}\frac{\textrm{Im}(w)}{|z-w|^2}$.
We prove in Supplemental Material {\color{red}S3}\,\cite{SupMat} that for a real Möbius transformation
\[
\Phi_t(z)=\frac{\alpha(t) z+\beta(t)}{\gamma(t) z+\delta(t)},
\qquad
\alpha\delta-\beta\gamma=1,
\]
the push-forward of the initial density by $\Phi_t$ is given by
\[
\rho(z,t)=(\Phi_{t\sharp} \rho_w)(z)
=
\frac{1}{\pi}
\frac{\textrm{Im}(\Phi_t(w))}{|z-\Phi_t(w)|^2}
=
\rho_{\Phi_t(w)}(z),
\]
More precisely, this identity is proved in the time-independent form $\Phi_\sharp\rho_w=\rho_{\Phi(w)}$ in 
Lemma~{\color{red}S3.3} of the Supplemental Material\,\cite{SupMat}. Therefore, the push-forward of a Cauchy-Lorentz density with parameter $w$ is another Cauchy-Lorentz density, with parameter $\Phi_t(w)$. In other words, the Cauchy-Lorentz family of densities is preserved by the push-forward action of the Möbius group.


The next question then becomes: is Cauchy-Lorentz the unique family of densities that is preserved under this push-forward action?

Our main result is a classification statement that is proved in Supplemental Material {\color{red}S4}\,\cite{SupMat}. Let $\mathcal P$ denote the set of continuous probability densities on $S^1$, and let $G=\PSL(2,\mathbb R)$ act on $\mathcal P$ by push-forward. 
We show that any connected two-dimensional  smooth submanifold $\mathcal M\subset\mathcal P$ (embedded, i.e., without self-intersection) that is invariant under $G$ is, under stereographic identification with the compactified real line $\widehat{\mathbb R}=\mathbb R\cup\{\infty\}$, exactly the Cauchy-Lorentz family.
The proof proceeds via the orbit-stabilizer theorem under a transitivity hypothesis, which turns out to be automatic in this setting: the action of $G$ on $\mathcal P$ admits no orbit of dimension $0$ or $1$, so any connected two-dimensional invariant submanifold of continuous densities is necessarily a single $G$-orbit (Corollary~{\color{red}S4.14}). The Cauchy-Lorentz family is thus the unique connected two-parameter family of continuous densities preserved by the Möbius flow.

\begin{figure}[t]
    \centering
    \includegraphics[width=\columnwidth]{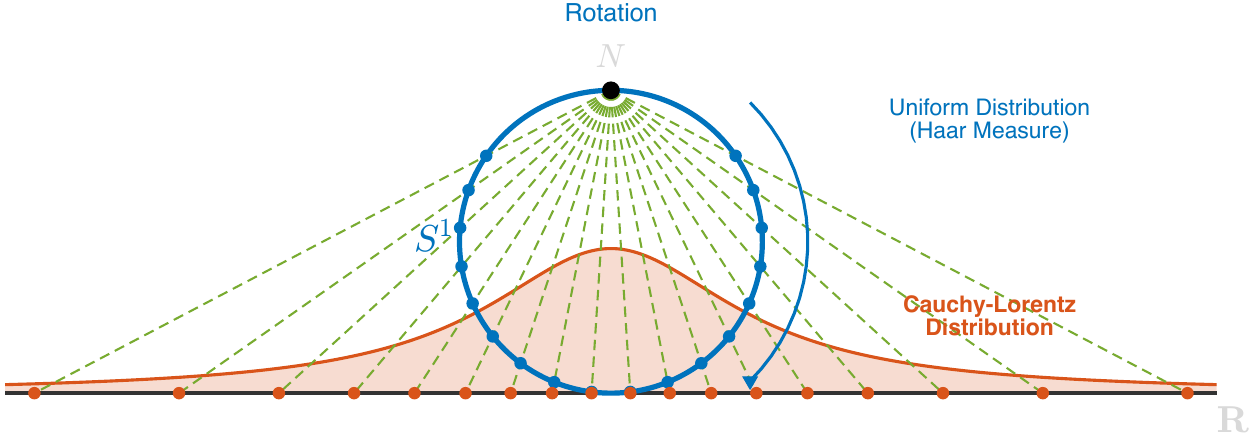}
    \caption{\textbf{Geometric origin of the Lorentzian Ansatz.} The macroscopic state is defined on the real line $\mathbb R$ (bottom). Via stereographic projection from the North Pole, the dynamics is represented on the compact circle $S^1$ (top). The unique rotation-invariant seed probability measure on $S^1$ is the Haar measure. Under stereographic projection, this seed becomes the standard Cauchy density on $\mathbb R$, and its full $\PSL(2,\mathbb R)$ orbit yields the Cauchy-Lorentz family.}
    \label{fig:geometry}
\end{figure}

The proof is geometric. It is convenient to move from the unbounded real line to the compact circle $S^1$ via the stereographic projection
\[
z=\tan(\theta/2)
\]
(Fig.~\ref{fig:geometry}). This identifies $\widehat{\mathbb R}$ with $S^1$ and conjugates the action of $\PSL(2,\mathbb R)$ on $\widehat{\mathbb R}$ to a M\"obius action on the circle, under which elliptic subgroups act as rigid rotations. 
With this compactified picture, the classification follows from an orbit-stabilizer argument (detailed in the Supplemental Material\,\cite{SupMat}): any density on a two-dimensional invariant manifold must be left unchanged by some one-parameter subgroup of $G$, and the question is which of these subgroups are compatible with a continuous, strictly positive density on $S^1$. As noted earlier, parabolic and hyperbolic subgroups generate flows with fixed points and are incompatible with such a density; only elliptic subgroups--acting as rigid rotations after a M\"obius change of coordinates--are admissible.

The only continuous probability density  that is invariant under such a rotation is the uniform Haar density
\[
g_H(\theta)=\frac{1}{2\pi},
\]
whose stereographic image is the standard Cauchy density
\[
\rho_0(z)=\frac{1}{\pi(1+z^2)}.
\]
Its full $G$-orbit--the set of all push-forwards $\{\Phi_\sharp\rho_0 : \Phi\in G\}$--is precisely the Cauchy-Lorentz family
\[
\rho_{x,y}(z)=\frac{1}{\pi}\frac{y}{(z-x)^2+y^2}.
\]

Supplemental Material~\cite{SupMat} provides the detailed functional setting, the explicit algebraic verification of the M\"obius covariance, and the complete proof.

This completes the dynamical picture: if the initial state $\rho_0$ is chosen from the Lorentzian orbit, the push-forward $(\Phi_t)_\sharp\rho_0$ remains within the same orbit for all time, with parameters $(x(t),y(t))$ evolving under the induced projective action,  and the Lorentzian orbit is the only one to exhibit this invariance.

Any other choice of initial distribution would therefore not be preserved. An interesting example is the Gaussian family of distributions. A simple way to see why Gaussian closures fail is to examine a representative non-affine Möbius map, such as the inversion $z\mapsto -1/z$. Under push-forward, the density acquires the Jacobian factor $|dz'/dz|=z^{-2}$. This is compatible with the algebraic tails of the Cauchy family, but not with the Gaussian family, whose exponential tails are not preserved under such projective distortion. Thus Gaussian profiles do not define an invariant family under Möbius push-forward.  Numerical simulations illustrating Lorentzian invariance versus Gaussian non-invariance under Riccati flow are provided in Fig.~{\color{red}S1} of Supplemental Material {\color{red}S5}~\cite{SupMat}. Our general uniqueness statement, however, follows from the orbit-stabilizer argument above and not from any particular example.

\paragraph{Discussion and Outlook.---}

Our results clarify the structural origin of exact mean-field reductions for coupled oscillators and spiking neurons. In the applied literature, the Cauchy-Lorentz profile is usually introduced as the \emph{Lorentzian Ansatz}, a terminology that naturally suggests a convenient closure hypothesis. The geometric picture developed here shows that, in Riccati-driven transport, its role is more rigid: under the Möbius push-forward symmetry induced by the microscopic flow, the Lorentzian family is the unique connected two-dimensional homogeneous invariant orbit of continuous densities.

It is important to position this statement correctly. The exceptional role of the Cauchy family under real linear fractional transformations is classical in probability theory and goes back, in different forms, to the characterization results of Knight and the preservation theory of Letac\,\cite{knight1976cauchy,letac1977cauchy}. The contribution of the present work is to recast that rigidity in the language relevant to exact mean-field reductions: Riccati transport, push-forward dynamics of probability densities, and invariant low-dimensional manifolds. In this sense, our paper does not merely restate a classical fact; it explains why the so-called Lorentzian Ansatz of OA/MPR-type theories is not an arbitrary guess, but the geometric orbit selected by the underlying projective symmetry.

Our result also complements the foundational work of Marvel, Mirollo, and Strogatz\,\cite{Marvel2009}, who established the invariance of the Poisson family under the M\"obius push-forward action and identified it as the orbit of the uniform measure on $S^1$. The classification theorem proved here closes the converse direction: no other connected two-dimensional family of continuous densities can be invariant under the same action. The Lorentzian Ansatz is therefore not one of several possible reductions, but the unique two-parameter reduction geometrically compatible with Riccati-driven transport.

Our derivation also clarifies the status of the macroscopic variables $(x,y)$. They are not merely convenient parameters or approximate moments, but geometric coordinates on the projective orbit generated from the uniform seed on the circle. This explains the ``unreasonable effectiveness'' of the OA and MPR reductions: they do not arise from an ad hoc closure imposed from outside the dynamics, but from restriction to the unique two-dimensional invariant orbit compatible with Riccati-induced Möbius transport.

Furthermore, this geometric viewpoint unifies the apparently different reduction schemes for phase oscillators (OA) and excitable neurons (MPR). At the level of state transport, the Lorentzian orbit on $\mathbb R$ and the Poisson-kernel orbit on $S^1$ are the same object written in different coordinates. The Supplemental Material shows that the corresponding order parameter evolves equivariantly under the same projective dynamics. The full MPR equations are then recovered once the standard Lorentzian heterogeneity closure is performed.

Our framework also clarifies two distinct mechanisms by which exact low-dimensional closure can fail. First, a family such as the Gaussian one is not invariant under Möbius push-forward, even when the underlying deterministic flow remains exactly Riccati/Möbius. In that case, the failure is one of closure, not of symmetry of the microscopic transport. Second, genuinely non-Riccati perturbations, such as diffusion or higher-order terms, break the projective structure of the transport operator and drive the dynamics away from the Lorentzian orbit. In both cases, exact two-parameter closure is lost, although for conceptually different reasons. 

This distinction suggests a natural diagnostic role for the Lorentzian profile in Riccati-driven systems. In regimes where the dynamics is expected to remain close to the Möbius-invariant orbit, deviations from the Cauchy-Lorentz form may signal either off-manifold initial preparation or genuine symmetry-breaking perturbations such as diffusion or non-Riccati corrections\,\cite{ratas2017symmetry,Tyulkina2018,klinshov2021reduction,goldobin2021reduction}. This provides a principled geometric benchmark for interpreting departures from exact low-dimensional closure in experimental or numerical data, for instance in arrays of Josephson junctions\,\cite{Watanabe1994} or in neuromorphic implementations of QIF-like dynamics.

In conclusion, the standard Cauchy law is the stereographic shadow of uniformity. More precisely, the Lorentzian family is the unique connected two-dimensional homogeneous invariant orbit of continuous densities under Möbius push-forward. This gives a geometric foundation for exact two-parameter reductions in Riccati-driven transport and clarifies, for the OA/MPR framework and its applications, why the Lorentzian closure is structural rather than merely heuristic.

\paragraph{Code Availability.---}
The source code used to generate the numerical results and figures presented in the Supplemental Material is openly available in the Inria GitLab repository at \url{https://gitlab.inria.fr/aistrosight/riccati-mobius-dynamics}.

\begin{acknowledgments}
This work was supported by the French National Research Agency (ANR) under Grant No.~\href{https://anr.fr/Project-ANR-21-CE16-0022}{ANR-21-CE16-0022}~(EngFlea).
\end{acknowledgments}

\bibliography{biblio_manuscript.bib}

\begin{thebibliography}{19}%
\makeatletter
\providecommand \@ifxundefined [1]{%
 \@ifx{#1\undefined}
}%
\providecommand \@ifnum [1]{%
 \ifnum #1\expandafter \@firstoftwo
 \else \expandafter \@secondoftwo
 \fi
}%
\providecommand \@ifx [1]{%
 \ifx #1\expandafter \@firstoftwo
 \else \expandafter \@secondoftwo
 \fi
}%
\providecommand \natexlab [1]{#1}%
\providecommand \enquote  [1]{``#1''}%
\providecommand \bibnamefont  [1]{#1}%
\providecommand \bibfnamefont [1]{#1}%
\providecommand \citenamefont [1]{#1}%
\providecommand \href@noop [0]{\@secondoftwo}%
\providecommand \href [0]{\begingroup \@sanitize@url \@href}%
\providecommand \@href[1]{\@@startlink{#1}\@@href}%
\providecommand \@@href[1]{\endgroup#1\@@endlink}%
\providecommand \@sanitize@url [0]{\catcode `\\12\catcode `\$12\catcode
  `\&12\catcode `\#12\catcode `\^12\catcode `\_12\catcode `\%12\relax}%
\providecommand \@@startlink[1]{}%
\providecommand \@@endlink[0]{}%
\providecommand \url  [0]{\begingroup\@sanitize@url \@url }%
\providecommand \@url [1]{\endgroup\@href {#1}{\urlprefix }}%
\providecommand \urlprefix  [0]{URL }%
\providecommand \Eprint [0]{\href }%
\providecommand \doibase [0]{https://doi.org/}%
\providecommand \selectlanguage [0]{\@gobble}%
\providecommand \bibinfo  [0]{\@secondoftwo}%
\providecommand \bibfield  [0]{\@secondoftwo}%
\providecommand \translation [1]{[#1]}%
\providecommand \BibitemOpen [0]{}%
\providecommand \bibitemStop [0]{}%
\providecommand \bibitemNoStop [0]{.\EOS\space}%
\providecommand \EOS [0]{\spacefactor3000\relax}%
\providecommand \BibitemShut  [1]{\csname bibitem#1\endcsname}%
\let\auto@bib@innerbib\@empty
\bibitem [{\citenamefont {Watanabe}\ and\ \citenamefont
  {Strogatz}(1993)}]{WatanabeStrogatz1993}%
  \BibitemOpen
  \bibfield  {author} {\bibinfo {author} {\bibfnamefont {S.}~\bibnamefont
  {Watanabe}}\ and\ \bibinfo {author} {\bibfnamefont {S.~H.}\ \bibnamefont
  {Strogatz}},\ }\bibfield  {title} {\bibinfo {title} {Integrability of a
  globally coupled oscillator array},\ }\href
  {https://doi.org/10.1103/PhysRevLett.70.2391} {\bibfield  {journal} {\bibinfo
   {journal} {Physical Review Letters}\ }\textbf {\bibinfo {volume} {70}},\
  \bibinfo {pages} {2391} (\bibinfo {year} {1993})}\BibitemShut {NoStop}%
\bibitem [{\citenamefont {Watanabe}\ and\ \citenamefont
  {Strogatz}(1994)}]{Watanabe1994}%
  \BibitemOpen
  \bibfield  {author} {\bibinfo {author} {\bibfnamefont {S.}~\bibnamefont
  {Watanabe}}\ and\ \bibinfo {author} {\bibfnamefont {S.~H.}\ \bibnamefont
  {Strogatz}},\ }\bibfield  {title} {\bibinfo {title} {Constants of motion for
  superconducting {J}osephson arrays},\ }\href
  {https://doi.org/10.1016/0167-2789(94)90196-1} {\bibfield  {journal}
  {\bibinfo  {journal} {Physica D: Nonlinear Phenomena}\ }\textbf {\bibinfo
  {volume} {74}},\ \bibinfo {pages} {197} (\bibinfo {year} {1994})}\BibitemShut
  {NoStop}%
\bibitem [{\citenamefont {Goebel}(1995)}]{Goebel1995}%
  \BibitemOpen
  \bibfield  {author} {\bibinfo {author} {\bibfnamefont {C.~J.}\ \bibnamefont
  {Goebel}},\ }\bibfield  {title} {\bibinfo {title} {On the oscillatory
  dynamics of the {Kuramoto} model},\ }\href
  {https://doi.org/10.1016/0167-2789(94)00166-6} {\bibfield  {journal}
  {\bibinfo  {journal} {Physica D: Nonlinear Phenomena}\ }\textbf {\bibinfo
  {volume} {80}},\ \bibinfo {pages} {18} (\bibinfo {year} {1995})}\BibitemShut
  {NoStop}%
\bibitem [{\citenamefont {Marvel}\ \emph {et~al.}(2009)\citenamefont {Marvel},
  \citenamefont {Mirollo},\ and\ \citenamefont {Strogatz}}]{Marvel2009}%
  \BibitemOpen
  \bibfield  {author} {\bibinfo {author} {\bibfnamefont {S.~A.}\ \bibnamefont
  {Marvel}}, \bibinfo {author} {\bibfnamefont {R.~E.}\ \bibnamefont
  {Mirollo}},\ and\ \bibinfo {author} {\bibfnamefont {S.~H.}\ \bibnamefont
  {Strogatz}},\ }\bibfield  {title} {\bibinfo {title} {Identical phase
  oscillators with global sinusoidal coupling evolve by {M}öbius group
  action},\ }\href {https://doi.org/10.1063/1.3247089} {\bibfield  {journal}
  {\bibinfo  {journal} {Chaos}\ }\textbf {\bibinfo {volume} {19}},\ \bibinfo
  {pages} {043104} (\bibinfo {year} {2009})}\BibitemShut {NoStop}%
\bibitem [{\citenamefont {Ott}\ and\ \citenamefont
  {Antonsen}(2008)}]{ott2008low}%
  \BibitemOpen
  \bibfield  {author} {\bibinfo {author} {\bibfnamefont {E.}~\bibnamefont
  {Ott}}\ and\ \bibinfo {author} {\bibfnamefont {T.~M.}\ \bibnamefont
  {Antonsen}},\ }\bibfield  {title} {\bibinfo {title} {Low dimensional behavior
  of large systems of globally coupled oscillators},\ }\href
  {https://doi.org/10.1063/1.2930766} {\bibfield  {journal} {\bibinfo
  {journal} {Chaos}\ }\textbf {\bibinfo {volume} {18}},\ \bibinfo {pages}
  {037113} (\bibinfo {year} {2008})}\BibitemShut {NoStop}%
\bibitem [{\citenamefont {Montbrió}\ \emph {et~al.}(2015)\citenamefont
  {Montbrió}, \citenamefont {Pazó},\ and\ \citenamefont
  {Roxin}}]{montbrio2015macroscopic}%
  \BibitemOpen
  \bibfield  {author} {\bibinfo {author} {\bibfnamefont {E.}~\bibnamefont
  {Montbrió}}, \bibinfo {author} {\bibfnamefont {D.}~\bibnamefont {Pazó}},\
  and\ \bibinfo {author} {\bibfnamefont {A.}~\bibnamefont {Roxin}},\ }\bibfield
   {title} {\bibinfo {title} {Macroscopic description for networks of spiking
  neurons},\ }\href {https://doi.org/10.1103/PhysRevX.5.021028} {\bibfield
  {journal} {\bibinfo  {journal} {Physical Review X}\ }\textbf {\bibinfo
  {volume} {5}},\ \bibinfo {pages} {021028} (\bibinfo {year}
  {2015})}\BibitemShut {NoStop}%
\bibitem [{\citenamefont {Cestnik}\ and\ \citenamefont
  {Martens}(2024)}]{cestnik2024ntegrability}%
  \BibitemOpen
  \bibfield  {author} {\bibinfo {author} {\bibfnamefont {R.}~\bibnamefont
  {Cestnik}}\ and\ \bibinfo {author} {\bibfnamefont {E.~A.}\ \bibnamefont
  {Martens}},\ }\bibfield  {title} {\bibinfo {title} {Integrability of a
  globally coupled complex {R}iccati array: {Q}uadratic integrate-and-fire
  neurons, phase oscillators, and all in between},\ }\href
  {https://doi.org/10.1103/PhysRevLett.132.057201} {\bibfield  {journal}
  {\bibinfo  {journal} {Phys. Rev. Lett.}\ }\textbf {\bibinfo {volume} {132}},\
  \bibinfo {pages} {057201} (\bibinfo {year} {2024})}\BibitemShut {NoStop}%
\bibitem [{\citenamefont {Paz{\'o}}\ and\ \citenamefont
  {Cestnik}(2025)}]{pazo2025low}%
  \BibitemOpen
  \bibfield  {author} {\bibinfo {author} {\bibfnamefont {D.}~\bibnamefont
  {Paz{\'o}}}\ and\ \bibinfo {author} {\bibfnamefont {R.}~\bibnamefont
  {Cestnik}},\ }\bibfield  {title} {\bibinfo {title} {Low-dimensional dynamics
  of globally coupled complex riccati equations: Exact firing-rate equations
  for spiking neurons with clustered substructure},\ }\href
  {https://doi.org/10.1103/PhysRevE.111.L052201} {\bibfield  {journal}
  {\bibinfo  {journal} {Physical Review E}\ }\textbf {\bibinfo {volume}
  {111}},\ \bibinfo {pages} {L052201} (\bibinfo {year} {2025})},\ \bibinfo
  {note}
  {[\href{https://arxiv.org/abs/2503.15537}{arXiv:2503.15537}]}\BibitemShut
  {NoStop}%
\bibitem [{\citenamefont {Kisil}(2012)}]{kisil2012geometry}%
  \BibitemOpen
  \bibfield  {author} {\bibinfo {author} {\bibfnamefont {V.~V.}\ \bibnamefont
  {Kisil}},\ }\href {https://doi.org/https://doi.org/10.1142/p835} {\emph
  {\bibinfo {title} {Geometry of {M}\"obius transformations}}}\ (\bibinfo
  {publisher} {World Scientific},\ \bibinfo {year} {2012})\BibitemShut
  {NoStop}%
\bibitem [{\citenamefont {Halmos}(1974)}]{Halmos1950}%
  \BibitemOpen
  \bibfield  {author} {\bibinfo {author} {\bibfnamefont {P.~R.}\ \bibnamefont
  {Halmos}},\ }\href {https://doi.org/10.1007/978-1-4684-9440-2} {\emph
  {\bibinfo {title} {Measure {T}heory}}},\ \bibinfo {series} {Graduate Texts in
  Mathematics}, Vol.~\bibinfo {volume} {18}\ (\bibinfo  {publisher}
  {Springer-Verlag},\ \bibinfo {address} {New York},\ \bibinfo {year} {1974})\
  \bibinfo {note} {see Chapter XI on Haar Measure}\BibitemShut {NoStop}%
\bibitem [{Sup()}]{SupMat}%
  \BibitemOpen
  \href@noop {} {}\bibinfo {note} {See Supplemental Material at [URL will be
  inserted by publisher] for the linearization of the Riccati equation, the
  explicit M\"obius covariance of the Cauchy--Lorentz family, the full proof of
  the uniqueness theorem, and numerical validation against Gaussian initial
  conditions.}\BibitemShut {Stop}%
\bibitem [{\citenamefont {Hille}(1976)}]{hille1976ordinary}%
  \BibitemOpen
  \bibfield  {author} {\bibinfo {author} {\bibfnamefont {E.}~\bibnamefont
  {Hille}},\ }\href {https://books.google.fr/books?id=IjemPwAACAAJ} {\emph
  {\bibinfo {title} {Ordinary Differential Equations in the Complex Domain}}},\
  A Wiley-Interscience publication\ (\bibinfo  {publisher}
  {Wiley-Interscience},\ \bibinfo {address} {New York},\ \bibinfo {year}
  {1976})\ \bibinfo {note} {see Chapter 4 for Riccati equations}\BibitemShut
  {NoStop}%
\bibitem [{\citenamefont {Villani}(2009)}]{villani2009optimal}%
  \BibitemOpen
  \bibfield  {author} {\bibinfo {author} {\bibfnamefont {C.}~\bibnamefont
  {Villani}},\ }\href {https://doi.org/10.1007/978-3-540-71050-9} {\emph
  {\bibinfo {title} {Optimal Transport: Old and New}}},\ \bibinfo {series}
  {Grundlehren der mathematischen Wissenschaften}, Vol.\ \bibinfo {volume}
  {338}\ (\bibinfo  {publisher} {Springer},\ \bibinfo {address} {Berlin,
  Heidelberg},\ \bibinfo {year} {2009})\ \bibinfo {note} {see Chapter 5 for the
  dynamical formulation of the push forward}\BibitemShut {NoStop}%
\bibitem [{\citenamefont {Knight}(1976)}]{knight1976cauchy}%
  \BibitemOpen
  \bibfield  {author} {\bibinfo {author} {\bibfnamefont {F.~B.}\ \bibnamefont
  {Knight}},\ }\bibfield  {title} {\bibinfo {title} {A characterization of the
  {C}auchy type},\ }\href
  {https://doi.org/https://doi.org/10.1090/S0002-9939-1976-0394803-6}
  {\bibfield  {journal} {\bibinfo  {journal} {Proceedings of the American
  Mathematical Society}\ }\textbf {\bibinfo {volume} {55}},\ \bibinfo {pages}
  {130} (\bibinfo {year} {1976})}\BibitemShut {NoStop}%
\bibitem [{\citenamefont {Letac}(1977)}]{letac1977cauchy}%
  \BibitemOpen
  \bibfield  {author} {\bibinfo {author} {\bibfnamefont {G.}~\bibnamefont
  {Letac}},\ }\bibfield  {title} {\bibinfo {title} {Which functions preserve
  {C}auchy laws?},\ }\href
  {https://doi.org/https://doi.org/10.1090/S0002-9939-1977-0584393-8}
  {\bibfield  {journal} {\bibinfo  {journal} {Proceedings of the American
  Mathematical Society}\ }\textbf {\bibinfo {volume} {67}},\ \bibinfo {pages}
  {277} (\bibinfo {year} {1977})}\BibitemShut {NoStop}%
\bibitem [{\citenamefont {Ratas}\ and\ \citenamefont
  {Pyragas}(2017)}]{ratas2017symmetry}%
  \BibitemOpen
  \bibfield  {author} {\bibinfo {author} {\bibfnamefont {I.}~\bibnamefont
  {Ratas}}\ and\ \bibinfo {author} {\bibfnamefont {K.}~\bibnamefont
  {Pyragas}},\ }\bibfield  {title} {\bibinfo {title} {Symmetry breaking in two
  interacting populations of quadratic integrate-and-fire neurons},\ }\href
  {https://doi.org/10.1103/PhysRevE.96.042212} {\bibfield  {journal} {\bibinfo
  {journal} {Physical Review E}\ }\textbf {\bibinfo {volume} {96}},\ \bibinfo
  {pages} {042212} (\bibinfo {year} {2017})}\BibitemShut {NoStop}%
\bibitem [{\citenamefont {Tyulkina}\ \emph {et~al.}(2018)\citenamefont
  {Tyulkina}, \citenamefont {Goldobin}, \citenamefont {Klimenko},\ and\
  \citenamefont {Pikovsky}}]{Tyulkina2018}%
  \BibitemOpen
  \bibfield  {author} {\bibinfo {author} {\bibfnamefont {I.~V.}\ \bibnamefont
  {Tyulkina}}, \bibinfo {author} {\bibfnamefont {D.~S.}\ \bibnamefont
  {Goldobin}}, \bibinfo {author} {\bibfnamefont {L.~S.}\ \bibnamefont
  {Klimenko}},\ and\ \bibinfo {author} {\bibfnamefont {A.}~\bibnamefont
  {Pikovsky}},\ }\bibfield  {title} {\bibinfo {title} {Dynamics of noisy
  oscillator populations beyond the {O}tt-{A}ntonsen ansatz},\ }\href
  {https://doi.org/10.1103/PhysRevLett.120.264101} {\bibfield  {journal}
  {\bibinfo  {journal} {Phys. Rev. Lett.}\ }\textbf {\bibinfo {volume} {120}},\
  \bibinfo {pages} {264101} (\bibinfo {year} {2018})}\BibitemShut {NoStop}%
\bibitem [{\citenamefont {Klinshov}\ \emph {et~al.}(2021)\citenamefont
  {Klinshov}, \citenamefont {Kirillov},\ and\ \citenamefont
  {Nekorkin}}]{klinshov2021reduction}%
  \BibitemOpen
  \bibfield  {author} {\bibinfo {author} {\bibfnamefont {V.}~\bibnamefont
  {Klinshov}}, \bibinfo {author} {\bibfnamefont {S.}~\bibnamefont {Kirillov}},\
  and\ \bibinfo {author} {\bibfnamefont {V.}~\bibnamefont {Nekorkin}},\
  }\bibfield  {title} {\bibinfo {title} {Reduction of the collective dynamics
  of neural populations with realistic forms of heterogeneity},\ }\href
  {https://doi.org/10.1103/PhysRevE.103.L040302} {\bibfield  {journal}
  {\bibinfo  {journal} {Physical Review E}\ }\textbf {\bibinfo {volume}
  {103}},\ \bibinfo {pages} {L040302} (\bibinfo {year} {2021})}\BibitemShut
  {NoStop}%
\bibitem [{\citenamefont {Goldobin}\ \emph {et~al.}(2021)\citenamefont
  {Goldobin}, \citenamefont {Di~Volo},\ and\ \citenamefont
  {Torcini}}]{goldobin2021reduction}%
  \BibitemOpen
  \bibfield  {author} {\bibinfo {author} {\bibfnamefont {D.~S.}\ \bibnamefont
  {Goldobin}}, \bibinfo {author} {\bibfnamefont {M.}~\bibnamefont {Di~Volo}},\
  and\ \bibinfo {author} {\bibfnamefont {A.}~\bibnamefont {Torcini}},\
  }\bibfield  {title} {\bibinfo {title} {Reduction methodology for fluctuation
  driven population dynamics},\ }\href
  {https://doi.org/10.1103/PhysRevLett.127.038301} {\bibfield  {journal}
  {\bibinfo  {journal} {Physical Review Letters}\ }\textbf {\bibinfo {volume}
  {127}},\ \bibinfo {pages} {038301} (\bibinfo {year} {2021})}\BibitemShut
  {NoStop}%
\end{thebibliography}%


\begin{thebibliography}{5}%
\makeatletter
\providecommand \@ifxundefined [1]{%
 \@ifx{#1\undefined}
}%
\providecommand \@ifnum [1]{%
 \ifnum #1\expandafter \@firstoftwo
 \else \expandafter \@secondoftwo
 \fi
}%
\providecommand \@ifx [1]{%
 \ifx #1\expandafter \@firstoftwo
 \else \expandafter \@secondoftwo
 \fi
}%
\providecommand \natexlab [1]{#1}%
\providecommand \enquote  [1]{``#1''}%
\providecommand \bibnamefont  [1]{#1}%
\providecommand \bibfnamefont [1]{#1}%
\providecommand \citenamefont [1]{#1}%
\providecommand \href@noop [0]{\@secondoftwo}%
\providecommand \href [0]{\begingroup \@sanitize@url \@href}%
\providecommand \@href[1]{\@@startlink{#1}\@@href}%
\providecommand \@@href[1]{\endgroup#1\@@endlink}%
\providecommand \@sanitize@url [0]{\catcode `\\12\catcode `\$12\catcode
  `\&12\catcode `\#12\catcode `\^12\catcode `\_12\catcode `\%12\relax}%
\providecommand \@@startlink[1]{}%
\providecommand \@@endlink[0]{}%
\providecommand \url  [0]{\begingroup\@sanitize@url \@url }%
\providecommand \@url [1]{\endgroup\@href {#1}{\urlprefix }}%
\providecommand \urlprefix  [0]{URL }%
\providecommand \Eprint [0]{\href }%
\providecommand \doibase [0]{https://doi.org/}%
\providecommand \selectlanguage [0]{\@gobble}%
\providecommand \bibinfo  [0]{\@secondoftwo}%
\providecommand \bibfield  [0]{\@secondoftwo}%
\providecommand \translation [1]{[#1]}%
\providecommand \BibitemOpen [0]{}%
\providecommand \bibitemStop [0]{}%
\providecommand \bibitemNoStop [0]{.\EOS\space}%
\providecommand \EOS [0]{\spacefactor3000\relax}%
\providecommand \BibitemShut  [1]{\csname bibitem#1\endcsname}%
\let\auto@bib@innerbib\@empty
\bibitem [{\citenamefont {Hille}(1976)}]{hille1976ordinary}%
  \BibitemOpen
  \bibfield  {author} {\bibinfo {author} {\bibfnamefont {E.}~\bibnamefont
  {Hille}},\ }\href {https://books.google.fr/books?id=IjemPwAACAAJ} {\emph
  {\bibinfo {title} {Ordinary Differential Equations in the Complex Domain}}},\
  A Wiley-Interscience publication\ (\bibinfo  {publisher}
  {Wiley-Interscience},\ \bibinfo {address} {New York},\ \bibinfo {year}
  {1976})\ \bibinfo {note} {see Chapter 4 for Riccati equations}\BibitemShut
  {NoStop}%
\bibitem [{\citenamefont {Hardy}(2025)}]{hardy2025invariance}%
  \BibitemOpen
  \bibfield  {author} {\bibinfo {author} {\bibfnamefont {M.}~\bibnamefont
  {Hardy}},\ }\bibfield  {title} {\bibinfo {title} {Invariance of the {C}auchy
  family under linear fractional transformations},\ }\href
  {https://doi.org/10.1080/00029890.2025.2459048} {\bibfield  {journal}
  {\bibinfo  {journal} {The American Mathematical Monthly}\ }\textbf {\bibinfo
  {volume} {132}},\ \bibinfo {pages} {453} (\bibinfo {year}
  {2025})}\BibitemShut {NoStop}%
\bibitem [{\citenamefont {Letac}(1977)}]{letac1977cauchy}%
  \BibitemOpen
  \bibfield  {author} {\bibinfo {author} {\bibfnamefont {G.}~\bibnamefont
  {Letac}},\ }\bibfield  {title} {\bibinfo {title} {Which functions preserve
  {C}auchy laws?},\ }\href
  {https://doi.org/https://doi.org/10.1090/S0002-9939-1977-0584393-8}
  {\bibfield  {journal} {\bibinfo  {journal} {Proceedings of the American
  Mathematical Society}\ }\textbf {\bibinfo {volume} {67}},\ \bibinfo {pages}
  {277} (\bibinfo {year} {1977})}\BibitemShut {NoStop}%
\bibitem [{\citenamefont {Knapp}(2002)}]{knapp2002lie}%
  \BibitemOpen
  \bibfield  {author} {\bibinfo {author} {\bibfnamefont {A.~W.}\ \bibnamefont
  {Knapp}},\ }\href@noop {} {\emph {\bibinfo {title} {Lie Groups Beyond an
  Introduction}}},\ \bibinfo {edition} {2nd}\ ed.,\ \bibinfo {series} {Progress
  in Mathematics}, Vol.\ \bibinfo {volume} {140}\ (\bibinfo  {publisher}
  {Birkh\"auser},\ \bibinfo {address} {Boston},\ \bibinfo {year}
  {2002})\BibitemShut {NoStop}%
\bibitem [{\citenamefont {Knight}(1976)}]{knight1976cauchy}%
  \BibitemOpen
  \bibfield  {author} {\bibinfo {author} {\bibfnamefont {F.~B.}\ \bibnamefont
  {Knight}},\ }\bibfield  {title} {\bibinfo {title} {A characterization of the
  {C}auchy type},\ }\href
  {https://doi.org/https://doi.org/10.1090/S0002-9939-1976-0394803-6}
  {\bibfield  {journal} {\bibinfo  {journal} {Proceedings of the American
  Mathematical Society}\ }\textbf {\bibinfo {volume} {55}},\ \bibinfo {pages}
  {130} (\bibinfo {year} {1976})}\BibitemShut {NoStop}%
\end{thebibliography}%

\end{document}


\title{Supplemental Material for: \\ Geometric Origin of Exact Mean-Field Reductions: Möbius Symmetry and the Lorentzian Ansatz}

\author{Hugues Berry}
\affiliation{AIstroSight, Inria Lyon Centre, Hospices Civils de Lyon, Université Claude Bernard Lyon 1, F-69603 Villeurbanne, France}

\author{Leonardo Trujillo}
\affiliation{AIstroSight, Inria Lyon Centre, Hospices Civils de Lyon, Université Claude Bernard Lyon 1, F-69603 Villeurbanne, France}

\date{\today}

\maketitle

\onecolumngrid

\tableofcontents

\section{Linearization of Riccati Dynamics via Projective Transformation}
\label{sec:linearization_riccati}

In the main text, we asserted that the nonlinear Riccati equation is structurally induced by a linear system in a higher-dimensional projective space~\cite{hille1976ordinary}. Here, we provide the explicit derivation.

Consider the scalar Riccati equation for the state variable $z(t) \in \mathbb{R}$:
\begin{equation}
    \dot{z} = a(t)z^2 + b(t)z + c(t).
    \label{eq:riccati_supp}
\end{equation}
We introduce the projective transformation
\[
z(t) = \frac{v_1(t)}{v_2(t)},
\]
where $\mathbf{v} = (v_1, v_2)^\top$ is a vector in $\mathbb{R}^2$. Differentiating $z$ with respect to time using the quotient rule yields:
\begin{equation}
    \dot{z}
    =
    \frac{\dot{v}_1 v_2 - v_1 \dot{v}_2}{v_2^2}
    =
    \frac{\dot{v}_1}{v_2}
    -
    \frac{v_1}{v_2}\frac{\dot{v}_2}{v_2}
    =
    \frac{\dot{v}_1}{v_2}
    -
    z\frac{\dot{v}_2}{v_2}.
    \label{eq:z_dot_quotient}
\end{equation}

We assume that the lifted vector $\mathbf{v}$ evolves according to a linear system
\[
\dot{\mathbf{v}} = \mathcal{H}(t)\mathbf{v},
\]
where
\begin{equation}
    \mathcal{H}(t)
    =
    \begin{pmatrix}
        A(t) & B(t) \\
        C(t) & D(t)
    \end{pmatrix}.
\end{equation}
The components therefore satisfy
\begin{align}
    \dot{v}_1 &= A v_1 + B v_2, \\
    \dot{v}_2 &= C v_1 + D v_2.
\end{align}
Substituting these relations into Eq.~\eqref{eq:z_dot_quotient}, and using $v_1/v_2=z$, we obtain
\begin{align}
    \dot{z}
    &=
    \frac{A v_1 + B v_2}{v_2}
    -
    z\left(\frac{C v_1 + D v_2}{v_2}\right)
    \nonumber\\
    &=
    A z + B - z(Cz + D)
    \nonumber\\
    &=
    -C z^2 + (A-D)z + B.
\end{align}
Comparing this expression with the original Riccati equation \eqref{eq:riccati_supp}, we identify the coefficients:
\begin{equation}
    -C = a,
    \qquad
    A-D = b,
    \qquad
    B = c.
    \label{eq:matching_conditions}
\end{equation}

This system is underdetermined: it gives three constraints for the four entries of $\mathcal H(t)$. This non-uniqueness reflects the fact that the projective variable
\[
z=\frac{v_1}{v_2}
\]
is insensitive to a common rescaling of the lifted vector $(v_1,v_2)^\top$.

\subsection{Gauge freedom and the trace-free representative}
\label{subsec:gauge_trace_free}

The remaining degree of freedom in the lifting is precisely the addition of a scalar multiple of the identity. Indeed, let $\lambda(t)$ be an arbitrary real-valued function and define
\[
\widetilde{\mathcal H}(t)=\mathcal H(t)+\lambda(t)I
=
\begin{pmatrix}
A+\lambda & B\\
C & D+\lambda
\end{pmatrix}.
\]
Then the projected dynamics becomes
\[
\dot z
=
-Cz^2+\bigl((A+\lambda)-(D+\lambda)\bigr)z+B
=
-Cz^2+(A-D)z+B,
\]
which is exactly the same Riccati equation as before. Therefore, the transformation
\[
\mathcal H \longmapsto \mathcal H+\lambda(t)I
\]
does not change the induced dynamics on the projective variable $z$.

Geometrically, this means that the lift is defined only up to a common rescaling of the vector $(v_1,v_2)^\top$. Since
\[
\frac{\mu(t)v_1(t)}{\mu(t)v_2(t)}=\frac{v_1(t)}{v_2(t)},
\]
the projective coordinate does not detect the radial degree of freedom in $\mathbb R^2$. The scalar term $\lambda(t)I$ generates precisely this irrelevant radial mode.

A natural gauge fixing consists in selecting the trace-free representative
\[
\operatorname{Tr}(\mathcal H)=A+D=0.
\]
This does not impose any additional dynamical restriction on the Riccati equation; it simply selects a canonical representative within the gauge-equivalence class of all linear lifts. Under this condition, one has $D=-A$, and the coefficient matching \eqref{eq:matching_conditions} yields the unique trace-free generator
\begin{equation}
    \mathcal{H}(t)
    =
    \begin{pmatrix}
        \dfrac{b(t)}{2} & c(t) \\
        -a(t) & -\dfrac{b(t)}{2}
    \end{pmatrix}.
    \label{eq:trace_free_generator}
\end{equation}

This gauge choice is especially natural because it places the lifted dynamics in the Lie algebra
\[
\mathfrak{sl}(2,\mathbb R),
\]
the space of trace-free real \(2\times2\) matrices. If $\mathbf M(t)$ denotes the fundamental solution matrix of the lifted system,
\[
\dot{\mathbf M}(t)=\mathcal H(t)\mathbf M(t),
\qquad
\mathbf M(0)=I,
\]
then Liouville's formula gives
\[
\det \mathbf M(t)
=
\exp\!\left(\int_0^t \operatorname{Tr}(\mathcal H(s))\,ds\right).
\]
Hence, under the trace-free condition,
\[
\det \mathbf M(t)=1,
\]
so that
\[
\mathbf M(t)\in SL(2,\mathbb R).
\]
Since projective dynamics identifies matrices differing by a nonzero scalar factor, this is precisely the natural linear setting underlying the action of
\[
PSL(2,\mathbb R)=SL(2,\mathbb R)/\{\pm I\}.
\]

We conclude that the condition $\operatorname{Tr}(\mathcal H)=0$ is not required for the existence of a linear lifting of the Riccati equation. Rather, it is a canonical gauge choice that removes the dynamically irrelevant scalar mode, isolates the projective content of the flow, and places the lifted dynamics naturally in the Lie algebra $\mathfrak{sl}(2,\mathbb R)$, whose associated group action generates the M\"obius transformations underlying the exact mean-field reduction.

\section{Stereographic correspondence between the Cauchy and Haar measures}
\label{sec:stereographic_isomorphism}

In the main text, we utilized the stereographic projection to map the Riccati flow on $\mathbb{R}$ to rigid rotations on $S^1$. Here, we provide the rigorous derivation of this isomorphism, relying on the trigonometric identity highlighted by Hardy~\cite{hardy2025invariance} and the measure-theoretic connection noted by Letac~\cite{letac1977cauchy}.

\subsection{Trigonometric form of elliptic Möbius transformations}

We aim to show that the elliptic subgroup of $\text{PSL}(2, \mathbb{R})$ acts on the angle $\theta$ as a pure translation (rotation).
Let $z = \tan(\theta/2)$. Consider a Möbius transformation represented by a matrix $\mathbf{M} \in \text{SL}(2, \mathbb{R})$. For the specific case of oscillatory dynamics (elliptic elements), the matrix is conjugate to a rotation matrix:
\begin{equation}
    \mathbf{M} = \begin{pmatrix} \cos(\phi/2) & -\sin(\phi/2) \\ \sin(\phi/2) & \cos(\phi/2) \end{pmatrix}.
\end{equation}
The action on $z$ is:
\begin{equation}
    z' = \frac{\cos(\phi/2) z - \sin(\phi/2)}{\sin(\phi/2) z + \cos(\phi/2)}.
    \label{eq:S10_elliptic_action}
\end{equation}
Substituting $z = \tan(\theta/2)$ into Eq.~\eqref{eq:S10_elliptic_action}, dividing numerator and denominator by $\cos(\phi/2)$, and using the tangent subtraction formula $\tan(A-B) = \frac{\tan A - \tan B}{1 + \tan A \tan B}$:
\begin{equation}
    \tan(\theta'/2) = \frac{\tan(\theta/2) - \tan(\phi/2)}{1 + \tan(\theta/2)\tan(\phi/2)} = \tan\left( \frac{\theta - \phi}{2} \right).
\end{equation}
This implies $\theta' = \theta - \phi \pmod{2\pi}$.
This derivation recovers, in the elliptic case, the trigonometric reduction emphasized by Hardy~\cite{hardy2025invariance}: after the angular parametrization \(z=\tan(\theta/2)\), the corresponding linear fractional transformation becomes a rigid phase shift on the circle.

\subsection{Stereographic projection of the Haar measure}
The standard Cauchy law admits a classical stereographic interpretation as the projection of the uniform distribution on the circle, a connection discussed explicitly by Letac~\cite{letac1977cauchy}.

Let $\mu$ be the Haar measure on $S^1$, with density $g(\theta) = \frac{1}{2\pi}$. We seek the density $\rho(z)$ on $\mathbb{R}$ induced by the map $z(\theta) = \tan(\theta/2)$.
The conservation of probability requires $\rho(z) |dz| = g(\theta) |d\theta|$.
First, compute the differential:
\begin{equation}
    dz = \frac{d}{d\theta} \left( \tan \frac{\theta}{2} \right) d\theta = \frac{1}{2} \sec^2 \left( \frac{\theta}{2} \right) d\theta = \frac{1}{2} \left( 1 + \tan^2 \frac{\theta}{2} \right) d\theta = \frac{1}{2} (1 + z^2) d\theta.
\end{equation}
Thus, the Jacobian of the inverse transformation is $|d\theta/dz| = \frac{2}{1+z^2}$.
Substituting this into the density equation:
\begin{equation}
    \rho(z) = g(\theta(z)) \left| \frac{d\theta}{dz} \right| = \frac{1}{2\pi} \cdot \frac{2}{1+z^2} = \frac{1}{\pi(1+z^2)}.
\end{equation}
This explicitly proves that the projection of the Haar measure is the standard Cauchy distribution. Any rotation $\theta \to \theta + \phi$ leaves $g(\theta)$ invariant, and thus leaves the functional form of $\rho(z)$ invariant (modulo the parameter change induced by the map if it were not a pure rotation in the $z$-frame).

\section{Explicit Invariance of the Cauchy-Lorentz Family}
\label{sec:explicit_invariance}

In this section we establish the precise covariance of the Cauchy-Lorentz family under the push-forward action of the M\"obius group. Since a general real M\"obius transformation is not a global diffeomorphism of $\mathbb R$ (it may have a pole in a given real chart), the natural global setting is the compactified line
\[
\widehat{\mathbb R}\simeq S^1.
\]
We therefore formulate the push-forward action first on the circle, and only afterwards pass to the real coordinate $z=\tan(\theta/2)$.

\subsection{Push-forward of densities on the circle}

Let $S^1=\mathbb R/2\pi\mathbb Z$, endowed with angular coordinate $\theta$ and reference measure $d\theta$.

\begin{lemma}[Push-forward of densities on the circle]
\label{lem:pushforward_circle}
Let $\Phi:S^1\to S^1$ be a $C^1$ diffeomorphism, and let $\mu_0$ be a probability measure absolutely continuous with respect to $d\theta$, with density $g_0\in C^0(S^1)$:
\[
d\mu_0(\theta)=g_0(\theta)\,d\theta.
\]
Then the push-forward measure
\[
\mu_1:=\Phi_\sharp \mu_0
\]
is absolutely continuous with respect to $d\theta$, with density
\[
g_1(\theta')
=
g_0(\Phi^{-1}(\theta'))
\left|
\frac{d}{d\theta'}\Phi^{-1}(\theta')
\right|.
\]
\end{lemma}

\begin{proof}
Let $\varphi\in C^0(S^1)$. By definition of push-forward,
\[
\int_{S^1}\varphi(\theta')\,d\mu_1(\theta')
=
\int_{S^1}\varphi(\Phi(\theta))\,d\mu_0(\theta)
=
\int_{S^1}\varphi(\Phi(\theta))\,g_0(\theta)\,d\theta.
\]
Performing the change of variables
\[
\theta=\Phi^{-1}(\theta')
\]
gives
\[
\int_{S^1}\varphi(\theta')\,
g_0(\Phi^{-1}(\theta'))
\left|
\frac{d}{d\theta'}\Phi^{-1}(\theta')
\right|
\,d\theta'.
\]
Since this holds for every test function $\varphi$, the density of $\mu_1$ is
\[
g_1(\theta')
=
g_0(\Phi^{-1}(\theta'))
\left|
\frac{d}{d\theta'}\Phi^{-1}(\theta')
\right|.
\]
\end{proof}

\begin{remark}[Real-coordinate chart]
\label{rem:real_chart_pushforward}
Via the stereographic map
\[
z=\tan\!\left(\frac{\theta}{2}\right),
\]
the same formula holds in any real coordinate chart on $\widehat{\mathbb R}$ away from the pole of that chart. Thus the familiar expression
\[
\rho_1(z')
=
\rho_0(\Phi^{-1}(z'))
\left|
\frac{d}{dz'}\Phi^{-1}(z')
\right|
\]
should be understood as a local coordinate representation of the global push-forward formula on $S^1$.
\end{remark}

\subsection{Cauchy-Lorentz family and complex parametrization}

We now work in the real chart $z\in\mathbb R$, and consider the Cauchy-Lorentz family
\begin{equation}
\rho_{x,y}(z)
=
\frac{1}{\pi}\frac{y}{(z-x)^2+y^2},
\qquad x\in\mathbb R,\quad y>0.
\label{eq:cauchy_family_real}
\end{equation}
It is convenient to encode the parameters by the point
\[
w=x+iy\in\mathbb H:=\{u+iv\in\mathbb C:\ v>0\}.
\]
Then \eqref{eq:cauchy_family_real} may be written as
\begin{equation}
\rho_w(z)
=
\frac{1}{\pi}\frac{\textrm{Im}(w)}{|z-w|^2}.
\label{eq:cauchy_family_complex}
\end{equation}

\subsection{M\"obius covariance}

\begin{lemma}[M\"obius covariance of the Cauchy family]
\label{lem:mobius_covariance_cauchy}
Let
\[
\Phi(z)=\frac{\alpha z+\beta}{\gamma z+\delta},
\qquad
\alpha,\beta,\gamma,\delta\in\mathbb R,
\qquad
\alpha\delta-\beta\gamma=1,
\]
be a real M\"obius transformation. Let $w\in\mathbb H$, and let $\rho_w$ be the density defined in
\eqref{eq:cauchy_family_complex}. Then, under the push-forward action of $\Phi$,
\[
\Phi_\sharp \rho_w = \rho_{\Phi(w)}.
\]
Equivalently, if $Z\sim \rho_w$, then
\[
\Phi(Z)\sim \rho_{\Phi(w)}.
\]
\end{lemma}

\begin{proof}
In a real coordinate chart where $\Phi$ is represented by the fractional linear map above, its inverse is
\[
\Phi^{-1}(z')
=
\frac{\delta z'-\beta}{-\gamma z'+\alpha},
\]
and
\[
\frac{d}{dz'}\Phi^{-1}(z')
=
\frac{1}{(-\gamma z'+\alpha)^2}.
\]
Hence, by Remark~\ref{rem:real_chart_pushforward},
\[
(\Phi_\sharp \rho_w)(z')
=
\rho_w(\Phi^{-1}(z'))
\left|
\frac{d}{dz'}\Phi^{-1}(z')
\right|.
\]
Substituting \eqref{eq:cauchy_family_complex} gives
\[
(\Phi_\sharp \rho_w)(z')
=
\frac{1}{\pi}
\frac{\textrm{Im}(w)}{|\Phi^{-1}(z')-w|^2}
\cdot
\frac{1}{|-\gamma z'+\alpha|^2}.
\]
Now
\[
\Phi^{-1}(z')-w
=
\frac{\delta z'-\beta}{-\gamma z'+\alpha}-w
=
\frac{(\delta+\gamma w)z'-(\beta+\alpha w)}{-\gamma z'+\alpha}.
\]
Therefore
\[
|\Phi^{-1}(z')-w|^2\,|-\gamma z'+\alpha|^2
=
|(\delta+\gamma w)z'-(\beta+\alpha w)|^2.
\]
Factoring out $|\gamma w+\delta|^2$, we obtain
\[
|(\delta+\gamma w)z'-(\beta+\alpha w)|^2
=
|\gamma w+\delta|^2
\left|
z'-\frac{\alpha w+\beta}{\gamma w+\delta}
\right|^2.
\]
Hence
\[
(\Phi_\sharp \rho_w)(z')
=
\frac{1}{\pi}
\frac{\textrm{Im}(w)}{|\gamma w+\delta|^2}
\frac{1}{|z'-\Phi(w)|^2}.
\]
Using the standard identity for the action of $\PSL(2,\mathbb R)$ on the upper half-plane,
\[
\textrm{Im}(\Phi(w))
=
\frac{\textrm{Im}(w)}{|\gamma w+\delta|^2},
\]
we conclude that
\[
(\Phi_\sharp \rho_w)(z')
=
\frac{1}{\pi}
\frac{\textrm{Im}(\Phi(w))}{|z'-\Phi(w)|^2}
=
\rho_{\Phi(w)}(z').
\]
The identity has been derived away from the pole of $\Phi^{-1}$ in the real chart, but both sides extend to continuous functions on $\widehat{\mathbb R}\simeq S^1$ and coincide on a dense open set; by continuity they coincide everywhere. This proves the claim.
\end{proof}

\begin{corollary}[Orbit of the standard Cauchy law]
\label{cor:orbit_standard_cauchy}
Let
\[
\rho_0(z)=\frac{1}{\pi(1+z^2)}=\rho_i(z),
\]
where $i\in\mathbb H$. Then
\[
G\cdot \rho_0
=
\left\{
\rho_{x,y}(z)=\frac{1}{\pi}\frac{y}{(z-x)^2+y^2}
:\ x\in\mathbb R,\ y>0
\right\},
\]
where $G=\PSL(2,\mathbb R)$ acts by push-forward.
\end{corollary}

\begin{proof}
By Lemma~\ref{lem:mobius_covariance_cauchy}, every push-forward of $\rho_i$ is again Cauchy:
\[
\Phi_\sharp \rho_i=\rho_{\Phi(i)}.
\]
Thus
\[
G\cdot \rho_0
\subset
\{\rho_{x,y}:x\in\mathbb R,\ y>0\}.
\]
Conversely, the action of $\PSL(2,\mathbb R)$ on the upper half-plane $\mathbb H$ is transitive. Therefore for every
\[
w=x+iy\in\mathbb H
\]
there exists $\Phi\in G$ such that
\[
\Phi(i)=w.
\]
Applying Lemma~\ref{lem:mobius_covariance_cauchy} again yields
\[
\Phi_\sharp \rho_i = \rho_{\Phi(i)}=\rho_w=\rho_{x,y}.
\]
Hence every Cauchy-Lorentz density belongs to the orbit of $\rho_0$, proving equality.
\end{proof}

\subsection{Equivariance of the macroscopic parameter}

The previous lemma implies that the full two-parameter Cauchy family is a single homogeneous orbit, naturally parametrized by
\[
w=x+iy\in\mathbb H.
\]
Therefore, if the density evolves by push-forward under a time-dependent M\"obius flow $\Phi_t\in \PSL(2,\mathbb R)$, then
\[
\rho_{w(0)}
\longmapsto
(\Phi_t)_\sharp \rho_{w(0)}
=
\rho_{\Phi_t(w(0))}.
\]
Thus the macroscopic parameter evolves by the same projective action:
\[
w(t)=\Phi_t(w(0)).
\]

\begin{corollary}[Riccati evolution of the order parameter]
\label{cor:riccati_parameter}
Let $\Phi_t$ be the projective flow generated by the real Riccati equation
\[
\dot z = a(t)z^2+b(t)z+c(t),
\qquad a,b,c\in\mathbb R.
\]
Then the corresponding complex parameter
\[
w(t)\in\mathbb H
\]
satisfies the same Riccati equation, now understood on $\mathbb H\subset\mathbb C$:
\[
\dot w = a(t)w^2+b(t)w+c(t).
\]
\end{corollary}

\begin{proof}
The flow $\Phi_t$ is generated by the same fractional linear action on $\mathbb R$ and on $\mathbb H$. Therefore the trajectory of the point $w(t)=\Phi_t(w(0))$ satisfies the same Riccati equation as any point transported by the projective flow.
\end{proof}

Writing
\[
w=x+iy,
\qquad y>0,
\]
and allowing, for generality, a complex effective drive
\[
c(t)=c_r(t)+i\,c_i(t),
\]
one obtains
\begin{align}
\dot x &= a(x^2-y^2)+bx+c_r,
\label{eq:xdot_general_complex}\\
\dot y &= 2axy+by+c_i.
\label{eq:ydot_general_complex}
\end{align}

\begin{remark}[Deterministic transport vs.\ heterogeneous closure]
\label{rem:deterministic_vs_heterogeneous}
For the deterministic Riccati transport of a fixed state density, the microscopic coefficients are real, hence
\[
c_i=0.
\]
In that case \eqref{eq:xdot_general_complex}-\eqref{eq:ydot_general_complex} reduce to
\begin{align}
\dot x &= a(x^2-y^2)+bx+c,
\\
\dot y &= 2axy+by.
\end{align}
This is the geometric evolution law of the Cauchy parameters under real Riccati transport.

By contrast, the full Montbri\'o-Paz\'o-Roxin closure requires an additional step: one must integrate over a Lorentzian distribution of excitabilities and evaluate the resulting contour integral at the pole in the upper half-plane. That step produces an effective imaginary contribution in the macroscopic drive.
\end{remark}

\paragraph*{Connection with the MPR variables.}
In the QIF setting, one writes
\[
w=v+i\pi r,
\]
where $v$ is the Lorentzian center and $r$ is the firing rate. If the effective drive after integration over Lorentzian heterogeneity is
\[
c_{\mathrm{eff}}(t)=\eta_0 + J s(t) + i\Delta,
\]
then \eqref{eq:xdot_general_complex}-\eqref{eq:ydot_general_complex} yield
\begin{align}
\dot v &= a(v^2-\pi^2r^2)+bv+\eta_0+Js(t),
\\
\dot r &= 2avr+br+\frac{\Delta}{\pi},
\end{align}
and for the canonical QIF choice
\[
a=1,\qquad b=0,
\]
one recovers the MPR equations
\begin{align}
\dot v &= v^2-\pi^2r^2+\eta_0+Js(t),
\\
\dot r &= 2vr+\frac{\Delta}{\pi}.
\end{align}

\begin{remark}[Logical role of this section]
The purpose of the present section is twofold:
\begin{enumerate}
    \item to prove the exact covariance of the Cauchy-Lorentz family under the M\"obius push-forward action;
    \item to show that, once restricted to this orbit, the macroscopic parameter evolves equivariantly under the same projective dynamics.
\end{enumerate}
This establishes the geometric backbone of the OA/MPR reductions. The uniqueness of this two-dimensional orbit is proved separately in Section~\ref{sec:uniqueness_symmetry}.
\end{remark}

\section{Uniqueness via Symmetry: The Main Theorem}
\label{sec:uniqueness_symmetry}

The ingredients recalled in Sections~\ref{sec:linearization_riccati}-\ref{sec:explicit_invariance} are classical in origin: the present purpose is not to re-establish their historical priority, but to assemble them into a geometric orbit-based formulation adapted to Riccati-driven transport and exact mean-field reductions.

In this section we give a rigorous classification theorem for two-dimensional invariant manifolds of probability densities under the push-forward action of the M\"obius group. The key point is to distinguish carefully between: (i) invariance of a single density, (ii) invariance of a family of densities, and (iii) the orbit structure induced by a transitive group action.

\subsection{Functional setting and geometric preliminaries}

We work on the unit circle
\[
S^1=\mathbb{R}/2\pi\mathbb{Z},
\]
endowed with its standard angular coordinate $\theta$ and reference measure $d\theta$. The normalized Haar probability measure on $S^1$ is
\[
d\mu_H(\theta)=\frac{d\theta}{2\pi}.
\]

\begin{definition}[Ambient space of continuous probability densities]
We define
\[
\mathcal P
:=
\left\{
g\in C^0(S^1)\ :\ g(\theta)\ge 0,\ \int_{S^1} g(\theta)\,d\theta =1
\right\}.
\]
Thus each $g\in\mathcal P$ determines a Borel probability measure
\[
d\mu_g(\theta)=g(\theta)\,d\theta
\]
absolutely continuous with respect to $d\theta$.
\end{definition}

Via the stereographic map
\[
z=\tan\!\left(\frac{\theta}{2}\right),
\]
we identify $S^1$ with the compactified line
\[
\widehat{\mathbb R}=\mathbb R\cup\{\infty\}.
\]
Accordingly, every density on $S^1$ determines a probability measure on $\widehat{\mathbb R}$, and conversely.

\begin{definition}[Push-forward action]
Let $G=\PSL(2,\mathbb R)$. Through the stereographic identification, each $\Phi\in G$ acts as a $C^\infty$ diffeomorphism of $S^1$. The induced push-forward action on $\mathcal P$ is defined by
\[
\Phi_\sharp(g\,d\theta):=(\Phi_\sharp g)\,d\theta,
\]
where
\[
(\Phi_\sharp g)(\theta')
=
g(\Phi^{-1}(\theta'))
\left|
\frac{d}{d\theta'}\Phi^{-1}(\theta')
\right|.
\]
\end{definition}

\begin{definition}[Invariant manifold]
A subset $\mathcal M\subset \mathcal P$ is called $G$-invariant if
\[
\Phi_\sharp \mathcal M \subset \mathcal M
\qquad \forall \Phi\in G.
\]
\end{definition}

\begin{definition}[Transitive action]
The action of $G$ on $\mathcal M$ is called transitive if for every $g_1,g_2\in\mathcal M$ there exists $\Phi\in G$ such that
\[
g_2=\Phi_\sharp g_1.
\]
Equivalently, $\mathcal M$ is a single $G$-orbit.
\end{definition}

\begin{assumption}[Smooth restricted action]
\label{ass:smooth_restricted_action}
Whenever $\mathcal M\subset \mathcal P$ is a finite-dimensional embedded submanifold, we assume that $\mathcal M$ carries a $C^1$ manifold structure (intrinsic, not inherited from any ambient Banach structure on $\mathcal P$) such that the restricted map
\[
G\times \mathcal M \to \mathcal M,
\qquad
(\Phi,g)\mapsto \Phi_\sharp g,
\]
is a $C^1$ action of the Lie group $G$ on $\mathcal M$. This is the minimal regularity required to apply the orbit-stabilizer theorem and the local inverse function theorem.
\end{assumption}

\subsection{Auxiliary lemmas}

\begin{lemma}[Stationary continuity equation for an invariant density]
\label{lem:stationary_continuity}
Let $(R_t)_{t\in\mathbb R}$ be a $C^1$ flow on $S^1$ generated by the vector field
\[
X(\theta)=v(\theta)\,\partial_\theta,
\qquad v\in C^1(S^1).
\]
Let $g\in C^0(S^1)$, $g\ge 0$, and assume that the probability measure
\[
d\mu(\theta)=g(\theta)\,d\theta
\]
is invariant under the flow:
\[
(R_t)_\sharp \mu = \mu
\qquad \forall t\in\mathbb R.
\]
Then
\[
\int_{S^1} v(\theta)\,\varphi'(\theta)\,g(\theta)\,d\theta =0
\qquad
\forall \varphi\in C^1(S^1).
\]
Equivalently,
\[
\partial_\theta\!\bigl(v(\theta)g(\theta)\bigr)=0
\]
in the sense of distributions. Since $vg\in C^0(S^1)$, it follows that
\[
v(\theta)g(\theta)=C
\]
for some constant $C\in\mathbb R$.
\end{lemma}

\begin{proof}
Let $\varphi\in C^1(S^1)$. By invariance of $\mu$,
\[
\int_{S^1}\varphi(\theta)\,g(\theta)\,d\theta
=
\int_{S^1}\varphi(R_t(\theta))\,g(\theta)\,d\theta
\qquad \forall t\in\mathbb R.
\]
Define
\[
F(t):=\int_{S^1}\varphi(R_t(\theta))\,g(\theta)\,d\theta.
\]
Since $R_t$ is a $C^1$ flow and $\varphi\in C^1$, the map $t\mapsto F(t)$ is differentiable, and by invariance it is constant. Hence
\[
0=F'(0)=\int_{S^1}\varphi'(\theta)\,v(\theta)\,g(\theta)\,d\theta.
\]
This is exactly the weak formulation of
\[
\partial_\theta(vg)=0.
\]
Because $vg\in C^0(S^1)$, a distributional derivative equal to zero implies that $vg$ is constant.
\end{proof}

\begin{lemma}[Homogeneous-space structure]
\label{lem:homogeneous_space}
Let $G$ be a Lie group acting transitively and $C^1$-smoothly on a connected $C^1$ manifold $M$. Fix $m_*\in M$, and let
\[
H:=\{g\in G:\ g\cdot m_*=m_*\}
\]
be the stabilizer of $m_*$. Then $H$ is a closed Lie subgroup of $G$, and
\[
M \simeq G/H
\]
as a $C^1$ homogeneous space. In particular,
\[
\dim M = \dim G - \dim H.
\]
\end{lemma}

\begin{proof}
This is the standard orbit-stabilizer theorem for smooth Lie group actions on manifolds.
\end{proof}

\begin{remark}[Classification of one-parameter subgroups of $\PSL(2,\mathbb R)$]
\label{rem:one_parameter_classification}
Every nontrivial connected one-parameter subgroup of $\PSL(2,\mathbb R)$ is, up to conjugacy, of exactly one of the following three types~\cite{knapp2002lie}:
\begin{itemize}
    \item \emph{elliptic}: conjugate to the rotation subgroup
    \[
    \left\{
    \begin{pmatrix}\cos t & -\sin t \\ \sin t & \cos t\end{pmatrix}
    : t\in\mathbb R
    \right\},
    \]
    which acts on $S^1$ without fixed points;
    \item \emph{parabolic}: conjugate to the upper-triangular unipotent subgroup
    \[
    \left\{
    \begin{pmatrix}1 & t \\ 0 & 1\end{pmatrix}
    : t\in\mathbb R
    \right\},
    \]
    which has one (double) fixed point on $S^1$;
    \item \emph{hyperbolic}: conjugate to the diagonal subgroup
    \[
    \left\{
    \begin{pmatrix}e^{t/2} & 0 \\ 0 & e^{-t/2}\end{pmatrix}
    : t\in\mathbb R
    \right\},
    \]
    which has two (simple) fixed points on $S^1$.
\end{itemize}
This classification follows from the Jordan decomposition in $\mathfrak{sl}(2,\mathbb R)$ via the sign of the discriminant of the generator.
\end{remark}

\begin{lemma}[Exclusion of parabolic and hyperbolic stabilizers]
\label{lem:exclude_nonelliptic}
Let $(R_t)_{t\in\mathbb R}$ be a nontrivial one-parameter subgroup of $\PSL(2,\mathbb R)$ acting on $S^1$. Assume that its generator
\[
X(\theta)=v(\theta)\,\partial_\theta
\]
has at least one zero on $S^1$. Then there exists no continuous probability density
\[
g\in C^0(S^1),\qquad g\ge 0,\qquad \int_{S^1}g(\theta)\,d\theta=1,
\]
such that
\[
(R_t)_\sharp(g\,d\theta)=g\,d\theta
\qquad \forall t\in\mathbb R.
\]
\end{lemma}

\begin{proof}
Assume by contradiction that such a density $g$ exists. By Lemma~\ref{lem:stationary_continuity},
\[
v(\theta)g(\theta)=C
\]
for some constant $C$.

Since $v$ has at least one zero, say at $\theta_0$, continuity gives
\[
C=v(\theta_0)g(\theta_0)=0.
\]
Hence
\[
v(\theta)g(\theta)=0
\qquad \forall \theta\in S^1.
\]
By Remark~\ref{rem:one_parameter_classification}, the zero set of a nontrivial parabolic or hyperbolic vector field is finite, so $v(\theta)\neq 0$ on a nonempty open dense subset of $S^1$.
Therefore
\[
g(\theta)=0
\]
on that open dense subset. By continuity, $g\equiv 0$ on all of $S^1$, contradicting
\[
\int_{S^1} g(\theta)\,d\theta =1.
\]
\end{proof}

\begin{lemma}[Elliptic subgroups are Möbius-conjugate to rigid rotations]
\label{lem:elliptic_conjugacy}
Let \(H\subset \PSL(2,\mathbb R)\) be a nontrivial one-parameter elliptic subgroup, and let \((R_t)_{t\in\mathbb R}\) denote its induced action on \(S^1\). Then there exists a diffeomorphism
\[
\Psi:S^1\to S^1
\]
which is itself induced by an element of \(\PSL(2,\mathbb R)\), and a constant \(\omega\neq 0\), such that
\[
\Psi\circ R_t\circ \Psi^{-1}(\theta)
=
\theta+\omega t
\qquad (\mathrm{mod}\ 2\pi)
\]
for all \(t\in\mathbb R\).
\end{lemma}

\begin{proof}
Via the Cayley transform, we identify \(\PSL(2,\mathbb R)\) with the group \(\PSU(1,1)\) of orientation-preserving Möbius transformations of the unit disk \(\mathbb D\). Under this identification, the subgroup \(H\) acts on \(\mathbb D\) by Möbius maps and on \(S^1=\partial\mathbb D\) by their boundary values.

Since \(H\) is elliptic, it has a unique fixed point \(p\in\mathbb D\). Because \(\PSU(1,1)\) acts transitively on \(\mathbb D\), there exists a Möbius transformation
\[
T\in \PSU(1,1)
\]
such that
\[
T(p)=0.
\]
Consider the conjugated subgroup
\[
\widetilde R_t := T\circ R_t\circ T^{-1}.
\]
Then \(\widetilde R_t\) fixes \(0\) for every \(t\).

Now the stabilizer of \(0\) in \(\PSU(1,1)\) is precisely the rotation group
\[
z\mapsto e^{i\omega t}z
\]
for some \(\omega\in\mathbb R\). Since \(H\) is nontrivial, one has \(\omega\neq 0\). Therefore
\[
\widetilde R_t(z)=e^{i\omega t}z
\qquad \forall z\in \mathbb D.
\]
Restricting to the boundary \(S^1\), and writing \(\Psi:=T|_{S^1}\), we obtain
\[
\Psi\circ R_t\circ \Psi^{-1}(\theta)=\theta+\omega t
\qquad (\mathrm{mod}\ 2\pi).
\]

Since \(T\) is a Möbius transformation in \(\PSU(1,1)\), its boundary action \(\Psi\) is induced by an element of \(\PSL(2,\mathbb R)\). Note that the two identifications of \(S^1\) used in this Supplemental Material--the stereographic projection \(z=\tan(\theta/2)\) of Sections~\ref{sec:stereographic_isomorphism}-\ref{sec:explicit_invariance}, and the Cayley transform used here--differ by an element of \(\PSL(2,\mathbb R)\), so the conjugation result is invariant under this change of chart. This proves the claim.
\end{proof}


\begin{lemma}[Rotation invariance forces uniformity]
\label{lem:rotation_uniformity}
Let $g\in C^0(S^1)$ and suppose that for some $\omega\neq 0$,
\[
g(\theta+\omega t)=g(\theta)
\qquad \forall \theta\in S^1,\ \forall t\in\mathbb R.
\]
Then $g$ is constant on $S^1$.
\end{lemma}

\begin{proof}
Since $\omega\neq 0$, the set
\[
\{\theta+\omega t \ (\mathrm{mod}\ 2\pi):\ t\in\mathbb R\}
\]
is all of $S^1$. Hence $g$ takes the same value at every point of the circle, and is therefore constant.
\end{proof}

\subsection{Main theorem}

\begin{theorem}[Uniqueness of the Lorentzian manifold]
\label{thm:uniqueness_lorentzian_manifold}
Let $\mathcal M\subset \mathcal P$ be a connected, two-dimensional, $C^1$ embedded submanifold. Assume that:
\begin{enumerate}
    \item $\mathcal M$ is $G$-invariant;
    \item the action of $G=\PSL(2,\mathbb R)$ on $\mathcal M$ is transitive;
    \item Assumption~\ref{ass:smooth_restricted_action} holds for $\mathcal M$.
\end{enumerate}
Then $\mathcal M$ is the unique two-dimensional homogeneous orbit of continuous probability densities under the push-forward action of $G$. Equivalently, under stereographic identification with $\widehat{\mathbb R}$, its image is exactly the Cauchy-Lorentz family
\[
\mathcal L
=
\left\{
\rho_{x,y}(z)=\frac{1}{\pi}\frac{y}{(z-x)^2+y^2}
:\ x\in\mathbb R,\ y>0
\right\}.
\]
\end{theorem}

\begin{proof}
Fix a reference density $g_*\in\mathcal M$, and define its stabilizer
\[
H:=\{\Phi\in G:\ \Phi_\sharp g_*=g_*\}.
\]
By transitivity and Assumption~\ref{ass:smooth_restricted_action}, Lemma~\ref{lem:homogeneous_space} applies, so
\[
\mathcal M \simeq G/H
\]
and
\[
\dim \mathcal M = \dim G - \dim H.
\]
Since
\[
\dim \mathcal M =2,
\qquad
\dim G = \dim \PSL(2,\mathbb R)=3,
\]
it follows that
\[
\dim H=1.
\]
Thus $H$ is a nontrivial one-parameter subgroup of $G$.

We now classify the possible types of $H$.

\medskip

\noindent
\textbf{Step 1: Exclusion of parabolic and hyperbolic stabilizers.}
If $H$ were parabolic or hyperbolic, then its induced flow on $S^1$ would have at least one fixed point, equivalently its generating vector field would have at least one zero. By Lemma~\ref{lem:exclude_nonelliptic}, no continuous probability density can then be invariant under $H$. This contradicts the defining property of the stabilizer,
\[
\Phi_\sharp g_* = g_*
\qquad \forall \Phi\in H.
\]
Therefore $H$ cannot be parabolic or hyperbolic.

\medskip

\noindent

\textbf{Step 2: Reduction to rigid rotations.}
Hence \(H\) must be elliptic. By Lemma~\ref{lem:elliptic_conjugacy}, there exists a diffeomorphism
\[
\Psi:S^1\to S^1
\]
induced by an element of \(G=\PSL(2,\mathbb R)\), and a nonzero constant \(\omega\), such that the conjugated action of \(H\) is the rigid rotation flow
\[
\Psi\circ R_t\circ \Psi^{-1}(\theta)
=
\theta+\omega t
\qquad (\mathrm{mod}\ 2\pi),
\]
where \((R_t)_{t\in\mathbb R}\) denotes the action of \(H\) on \(S^1\).

Define
\[
\widetilde g_*:=\Psi_\sharp g_*.
\]
We claim that \(\widetilde g_*\) is invariant under the rigid rotation flow
\[
\widetilde R_t:=\Psi\circ R_t\circ \Psi^{-1}.
\]
Indeed, using the functoriality of push-forward, we have
\[
(\widetilde R_t)_\sharp \widetilde g_*
=
(\Psi\circ R_t\circ \Psi^{-1})_\sharp (\Psi_\sharp g_*)
=
\Psi_\sharp\bigl((R_t)_\sharp g_*\bigr).
\]
Since \(R_t\in H\) and \(H\) is the stabilizer of \(g_*\), one has
\[
(R_t)_\sharp g_*=g_*.
\]
Therefore
\[
(\widetilde R_t)_\sharp \widetilde g_*
=
\Psi_\sharp g_*
=
\widetilde g_*.
\]
Thus \(\widetilde g_*\) is invariant under the nontrivial rigid rotation flow
\[
\theta\mapsto \theta+\omega t
\qquad (\mathrm{mod}\ 2\pi).
\]

By Lemma~\ref{lem:rotation_uniformity}, any continuous probability density invariant under such a flow must be constant. Since \(\widetilde g_*\) is normalized, it follows that
\[
\widetilde g_*=g_H:=\frac{1}{2\pi}.
\]
Hence
\[
g_*=\Psi^{-1}_\sharp g_H.
\]
Since \(\Psi^{-1}\) is induced by an element of \(G\), we conclude that
\[
g_*\in G\cdot g_H.
\]

\medskip

\noindent
\textbf{Step 3: Identification of the orbit.}
By Step 2, the reference density \(g_*\) belongs to the \(G\)-orbit of the Haar density
\[
g_H(\theta)=\frac{1}{2\pi}.
\]
Since \(\mathcal M\) is a single \(G\)-orbit through \(g_*\), it follows that
\[
\mathcal M = G\cdot g_* = G\cdot g_H.
\]

Let
\[
\Sigma:S^1\to \widehat{\mathbb R}
\]
denote the stereographic projection, and let \(\Sigma_\sharp\) be the induced push-forward on probability densities. By construction, \(\Sigma\) intertwines the \(G\)-action on \(S^1\) with the projective action on \(\widehat{\mathbb R}\), namely
\[
\Sigma_\sharp(\Phi_\sharp g)=\Phi_\sharp(\Sigma_\sharp g),
\qquad
\forall \Phi\in G,\ \forall g\in\mathcal P.
\]

Now define
\[
\rho_0:=\Sigma_\sharp g_H.
\]
Since
\[
\Sigma(\theta)=z=\tan\!\left(\frac{\theta}{2}\right),
\qquad
\theta=2\arctan z,
\]
the change-of-variables formula gives
\[
\rho_0(z)\,dz = g_H(\theta)\,d\theta,
\]
hence
\[
\rho_0(z)=g_H(\theta(z))\left|\frac{d\theta}{dz}\right|.
\]
Using
\[
\frac{d\theta}{dz}
=
\frac{d}{dz}\bigl(2\arctan z\bigr)
=
\frac{2}{1+z^2},
\]
we obtain
\[
\rho_0(z)
=
\frac{1}{2\pi}\cdot\frac{2}{1+z^2}
=
\frac{1}{\pi(1+z^2)}.
\]
Thus the stereographic image of the uniform density on \(S^1\) is precisely the standard Cauchy density on \(\widehat{\mathbb R}\).

Applying \(\Sigma_\sharp\) to the orbit identity \(\mathcal M=G\cdot g_H\), and using equivariance, yields
\[
\Sigma_\sharp \mathcal M
=
\Sigma_\sharp(G\cdot g_H)
=
G\cdot \Sigma_\sharp g_H
=
G\cdot \rho_0.
\]
By Corollary~\ref{cor:orbit_standard_cauchy},
\[
G\cdot \rho_0=\mathcal L.
\]
Therefore the stereographic image of \(\mathcal M\) is exactly the Cauchy-Lorentz family. This proves the theorem.

\end{proof}

\subsection{Absence of lower-dimensional orbits and transitivity for free}
\label{subsec:no_low_dim_orbits}

The classification theorem assumes that the action of $G$ on $\mathcal M$ is transitive. We now show that, for the action of $G=\PSL(2,\mathbb R)$ on the space of continuous probability densities $\mathcal P$, this hypothesis is automatic for any connected two-dimensional invariant submanifold: there are no orbits of dimension $0$ or $1$, hence a connected $2$-dimensional invariant submanifold is necessarily a single orbit.

\begin{proposition}[Absence of low-dimensional orbits]
\label{prop:no_low_dim_orbits}
Let $G=\PSL(2,\mathbb R)$ act on $\mathcal P$ by push-forward. Then no $g\in\mathcal P$ has stabilizer of dimension $\ge 2$. Equivalently, $G$ has no orbit of dimension $0$ or $1$ in $\mathcal P$.
\end{proposition}

\begin{proof}
The connected closed subgroups of $\PSL(2,\mathbb R)$ are classified up to conjugacy~\cite{knapp2002lie}:
\begin{itemize}
    \item dimension $3$: the full group $G$;
    \item dimension $2$: the Borel subgroup $B$, conjugate to the group of upper-triangular matrices modulo $\{\pm I\}$, which is the semidirect product of a hyperbolic and a parabolic one-parameter subgroup;
    \item dimension $1$: the elliptic, parabolic, and hyperbolic one-parameter subgroups;
    \item dimension $0$: discrete subgroups.
\end{itemize}
In particular, every connected closed subgroup of $G$ of dimension $\ge 2$ contains a nontrivial parabolic one-parameter subgroup.

Suppose, for contradiction, that $g\in\mathcal P$ has stabilizer $H$ with $\dim H\ge 2$. The connected component of the identity $H^0$ is a closed connected subgroup of $G$ of dimension $\ge 2$, hence contains a nontrivial parabolic one-parameter subgroup $H'\subset H^0\subset H$. Since $g$ is fixed by all of $H$, it is in particular fixed by $H'$. But by Lemma~\ref{lem:exclude_nonelliptic}, no continuous probability density on $S^1$ can be invariant under a nontrivial parabolic flow. This contradicts $g\in\mathcal P$.

Hence every $g\in\mathcal P$ has stabilizer of dimension $\le 1$, and every $G$-orbit has dimension $\ge 2$.
\end{proof}

\begin{corollary}[Transitivity is automatic for connected two-dimensional invariant submanifolds]
\label{cor:transitivity_automatic}
Let $\mathcal M\subset \mathcal P$ be a connected, $C^1$ embedded, $G$-invariant submanifold of dimension $2$, satisfying Assumption~\ref{ass:smooth_restricted_action}. Then the action of $G$ on $\mathcal M$ is transitive, and Theorem~\ref{thm:uniqueness_lorentzian_manifold} applies without the transitivity hypothesis. In particular, under stereographic identification, $\mathcal M$ is exactly the Cauchy-Lorentz family.
\end{corollary}

\begin{proof}
Let $g_*\in\mathcal M$ and let $\mathcal O:=G\cdot g_*$ be its $G$-orbit. By Proposition~\ref{prop:no_low_dim_orbits}, $\dim \mathcal O\ge 2$. On the other hand, since $\mathcal M$ is $G$-invariant, $\mathcal O\subset\mathcal M$, hence $\dim\mathcal O\le \dim\mathcal M=2$. Therefore $\dim\mathcal O=2$.

By Assumption~\ref{ass:smooth_restricted_action}, the orbit map $\Phi\mapsto \Phi_\sharp g_*$ from $G$ to $\mathcal M$ is $C^1$, so $\mathcal O$ is an immersed $C^1$ submanifold of $\mathcal M$. Since $\mathcal O$ and $\mathcal M$ have the same dimension, the inclusion $\mathcal O\hookrightarrow\mathcal M$ is a local diffeomorphism (by the inverse function theorem applied at each point), and hence $\mathcal O$ is open in $\mathcal M$. The same argument applied to any $g\in\mathcal M$ shows that every $G$-orbit in $\mathcal M$ is open. Since the orbits partition $\mathcal M$ into disjoint open sets, and $\mathcal M$ is connected, there can be only one orbit: $\mathcal M=\mathcal O$. Hence the action of $G$ on $\mathcal M$ is transitive, and Theorem~\ref{thm:uniqueness_lorentzian_manifold} applies, yielding the conclusion.
\end{proof}

\section{Numerical validation}
\label{sec:numerical_validation}

To provide a concrete verification of the uniqueness principle derived in the main text, we performed numerical simulations comparing the evolution of two distinct probability densities under the same microscopic Riccati dynamics. This Monte Carlo experiment illustrates the practical consequences of the projective rigidity established in the main text, and refutes the common heuristic assumption that the specific functional form of a unimodal (``bell-shaped'') distribution is irrelevant for the mean-field dynamics. The Python source code is openly available at \url{https://gitlab.inria.fr/aistrosight/riccati-mobius-dynamics}.

\subsection{Experimental design}

\paragraph{Population initialization.}
We initialized two statistical ensembles, each consisting of $N = 10^5$ independent units.
\begin{itemize}
    \item \textbf{Ensemble A (Lorentzian):} Samples drawn from a standard Cauchy-Lorentz distribution $\rho_L(z) = \frac{1}{\pi(1+z^2)}$.
    \item \textbf{Ensemble B (Gaussian):} Samples drawn from a standard normal distribution $\rho_G(z) = \frac{1}{\sqrt{2\pi}} e^{-z^2/2}$.
\end{itemize}
Both distributions share the same location parameter (mode at $z=0$) and comparable scale parameters to ensure a fair visual comparison of the initial states.

\paragraph{Dynamical protocol.}
The units were subjected to the canonical Riccati flow:
\begin{equation}
    \dot{z} = z^2 + 1.
    \label{eq:riccati_flow}
\end{equation}
This specific dynamics was chosen because it corresponds to a rigid rotation of the angle $\theta$ on the unit circle ($\dot{\theta} = 2$). This is the most stringent test for structural stability because it mixes the ``center'' of the distribution with its ``tails'' as the distribution rotates through infinity in the projective space.

\paragraph{Exact integration method.}
A critical aspect of this protocol is the handling of the finite-time singularity (blow-up) inherent to Eq.~\eqref{eq:riccati_flow}. Standard numerical integrators (e.g., Runge-Kutta) introduce discretization errors that diverge as $z \to \infty$. To eliminate numerical artifacts and isolate the purely geometric effects, we employed the exact analytic propagator (the Möbius map) for the time evolution. For a time step $t$, the state evolves as:
\begin{equation}
    z(t) = \tan(t + \arctan(z_0)) = \frac{z_0 + \tan(t)}{1 - z_0 \tan(t)}.
\end{equation}
This map handles the topology of the extended real line $\widehat{\mathbb{R}}$ exactly, treating the passage through infinity as a smooth rotation. We set the evolution time to $t=0.8$, sufficient to produce a significant rotation/deformation of the initial density.

\subsection{Results}

\begin{figure}[h]
    \centering
    \includegraphics[width=0.8\textwidth]{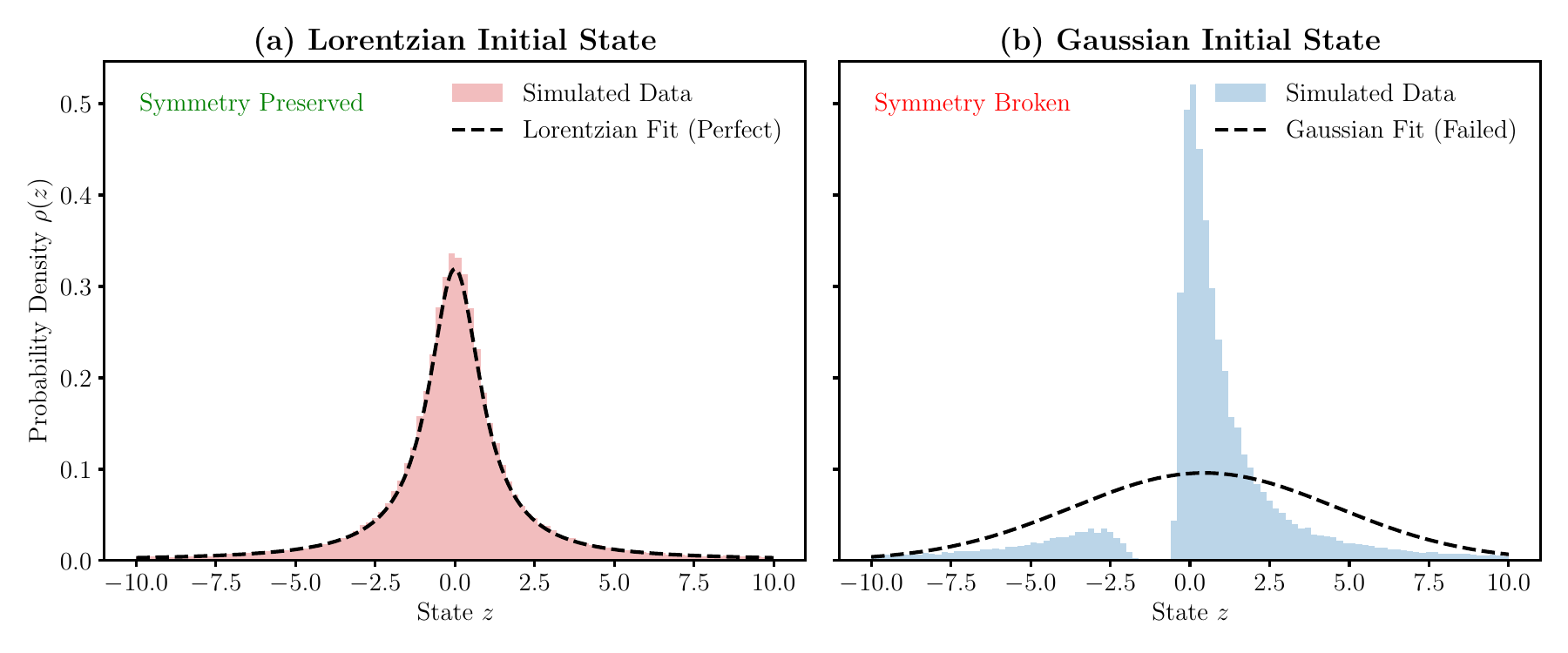}
    \caption{\textbf{Numerical validation of the uniqueness theorem.} We subjected two populations of $N=10^5$ units to the same microscopic Riccati dynamics ($\dot{z}=z^2+1$) starting from different initial distributions. \textbf{(a)} An initially Lorentzian population evolves into another perfect Lorentzian profile. \textbf{(b)} An initially Gaussian population is structurally deformed; the flow introduces heavy tails that a Gaussian fit fails to capture. This provides a heuristic illustration of the rigidity behind Knight's characterization theorem~\cite{knight1976cauchy}: among the standard low-dimensional candidates, the Cauchy type is the one compatible with projective invariance.}
    \label{fig:sim_validation}
\end{figure}

The post-evolution distributions were analyzed by attempting to fit them with their original functional families using Maximum Likelihood Estimation (MLE). The results are displayed in Fig.~\ref{fig:sim_validation}.

\begin{itemize}
    \item \textbf{Lorentzian invariance (Fig.~\ref{fig:sim_validation}a):} The evolved Ensemble A is perfectly fitted by a new Cauchy-Lorentz density (dashed line). The distribution has shifted and rescaled, but its functional form remains exact. This empirically confirms that the Lorentzian family is an invariant manifold of the dynamics.
    \item \textbf{Gaussian breakdown (Fig.~\ref{fig:sim_validation}b):} The evolved Ensemble B exhibits a catastrophic breakdown of its functional form. The resulting distribution is skewed and possesses heavy tails ($\sim z^{-2}$) induced by the Jacobian of the Möbius map. A Gaussian fit (dashed line) completely fails to capture the data, demonstrating that the Gaussian family is not invariant under the projective geometry of the system.
\end{itemize}

\paragraph*{Interpretation.}
This numerical experiment highlights a critical physical insight often overlooked in mean-field modeling: \emph{topology trumps shape}. While a Gaussian and a Lorentzian may look similar to the eye (both are bell-shaped), their algebraic properties under projective transformations are fundamentally different. The Möbius flow $\Phi_t(z)$ introduces a Jacobian factor $|d\Phi^{-1}/dz| \sim z^{-2}$ at large $z$. The polynomial tails of the Cauchy distribution ($\sim z^{-2}$) absorb this factor exactly, preserving the distribution type. The exponential tails of the Gaussian ($\sim e^{-z^2}$), however, cannot compensate for this algebraic scaling, leading to the immediate destruction of the distribution's form. Thus, Fig.~\ref{fig:sim_validation}(b) illustrates, at the level of a concrete Riccati flow, the kind of rigidity captured abstractly by Knight's characterization theorem~\cite{knight1976cauchy}.

\subsection{Source code}

For reproducibility, we provide below the Python script used to implement this protocol and generate Fig.~\ref{fig:sim_validation}.

\begin{lstlisting}[language=Python, caption=Python script for numerical validation of uniqueness.]
import numpy as np
import matplotlib.pyplot as plt
from scipy.stats import cauchy, norm

# Configuration for publication-quality plots
plt.rcParams.update({
    "text.usetex": True,
    "font.family": "serif",
    "font.serif": ["Computer Modern Roman"],
    "font.size": 14,
    "axes.linewidth": 1.5,
    "xtick.major.width": 1.5,
    "ytick.major.width": 1.5,
    "legend.frameon": False
})

def mobius_transform(z, t):
    """
    Applies the exact flow of the Riccati equation: dz/dt = z^2 + 1
    Analytical solution: z(t) = tan(t + arctan(z0))
    Moebius map form: z(t) = (z0 + tan(t)) / (1 - z0*tan(t))
    This avoids numerical integration errors near singularities.
    """
    return (z + np.tan(t)) / (1 - z * np.tan(t))

def run_simulation():
    # --- EXPERIMENTAL PARAMETERS ---
    N = 100000       # Sample size for Monte Carlo
    t_final = 0.8    # Evolution time (rotation angle)
    
    # Initial distribution parameters
    loc, scale = 0.0, 1.0 

    # --- 1. INITIALIZATION ---
    z_cauchy_0 = cauchy.rvs(loc=loc, scale=scale, size=N)
    z_gauss_0 = norm.rvs(loc=loc, scale=scale, size=N)

    # --- 2. DYNAMICS (EXACT PROPAGATOR) ---
    z_cauchy_t = mobius_transform(z_cauchy_0, t_final)
    z_gauss_t = mobius_transform(z_gauss_0, t_final)

    # --- 3. ANALYSIS AND PLOTTING ---
    fig, axes = plt.subplots(1, 2, figsize=(12, 5), sharey=True)
    plot_range = (-10, 10)
    bins = 100

    # --- LEFT PANEL: LORENTZIAN STABILITY ---
    ax = axes[0]
    ax.hist(z_cauchy_t, bins=bins, range=plot_range, density=True, 
            color='#d62728', alpha=0.3, label='Simulated Data')
    params_c = cauchy.fit(z_cauchy_t)
    x_grid = np.linspace(*plot_range, 1000)
    pdf_c = cauchy.pdf(x_grid, *params_c)
    ax.plot(x_grid, pdf_c, 'k--', linewidth=2, label=r'Lorentzian Fit (Perfect)')
    ax.set_title(r'\textbf{(a) Lorentzian Initial State}')
    ax.set_xlabel(r'State $z$')
    ax.set_ylabel(r'Probability Density $\rho(z)$')
    ax.legend(loc='upper right')
    ax.text(0.05, 0.9, r"Symmetry Preserved", transform=ax.transAxes, color='green', fontweight='bold')

    # --- RIGHT PANEL: GAUSSIAN INSTABILITY ---
    ax = axes[1]
    ax.hist(z_gauss_t, bins=bins, range=plot_range, density=True, 
            color='#1f77b4', alpha=0.3, label='Simulated Data')
    params_g = norm.fit(z_gauss_t[np.abs(z_gauss_t)<20]) 
    pdf_g = norm.pdf(x_grid, *params_g)
    ax.plot(x_grid, pdf_g, 'k--', linewidth=2, label=r'Gaussian Fit (Failed)')
    ax.set_title(r'\textbf{(b) Gaussian Initial State}')
    ax.set_xlabel(r'State $z$')
    ax.legend(loc='upper right')
    ax.text(0.05, 0.9, r"Symmetry Broken", transform=ax.transAxes, color='red', fontweight='bold')

    plt.tight_layout()
    plt.savefig('simulation_uniqueness.pdf', dpi=300)

if __name__ == "__main__":
    run_simulation()
\end{lstlisting}

\bibliography{biblio_SM.bib}